# Exogenous Metal Cations in the Synthesis of CsPbBr$_3$ Nanocrystals and their Interplay with Tertiary Amines

Zhanzhao Li[1], Luca Goldoni[2], Ye Wu[1], Muhammad Imran[1,†], Yurii P. Ivanov[3], Giorgio Divitini[3], Juliette Zito[1], Iyyappa Rajan Panneerselvam[1], Dmitry Baranov[1,4], Ivan Infante[5,6]*, Luca De Trizio[2]*, Liberato Manna[1]*

[1] Nanochemistry, Istituto Italiano di Tecnologia, Via Morego 30, Genova, Italy

[2] Chemistry Facility, Istituto Italiano di Tecnologia, Via Morego 30, Genova, Italy

[3] Electron Spectroscopy and Nanoscopy, Istituto Italiano di Tecnologia, Via Morego 30, Genova, Italy

[4] Division of Chemical Physics, Department of Chemistry, Lund University, P.O. Box, 124, SE-221 00 Lund, Sweden

[5] BCMaterials, Basque Center for Materials, Applications, and Nanostructures, UPV/EHU Science Park, Leioa 48940, Spain

[6] Ikerbasque Basque Foundation for Science Bilbao 48009, Spain

**ABSTRACT:** Current syntheses of CsPbBr$_3$ halide perovskite nanocrystals (NCs) rely on over-stoichiometric amounts of Pb$^{2+}$ precursors, resulting in unreacted lead ions at the end of the process. In our synthesis scheme of CsPbBr$_3$ NCs we replaced excess Pb$^{2+}$ with different exogenous metal cations (M) and investigated their effect on the synthesis products. These cations can be divided into two groups: group 1 delivers monodisperse CsPbBr$_3$ cubes capped with oleate species (as for the case when Pb$^{2+}$ is used in excess) and with photoluminescence quantum yield (PLQY) as high as 90% with some cations (for example with M= In$^{3+}$); group 2 yields irregularly shaped CsPbBr$_3$ NCs with broad size distributions. In both cases, the addition of a tertiary ammonium cation (didodecylmethyl ammonium, DDMA$^+$) during the synthesis, after the nucleation of the NCs, reshapes the NCs to monodisperse truncated cubes. Such NCs feature a mixed oleate/DDMA$^+$ surface termination with PLQY values up to 90%. For group 1 cations, this happens only if the ammonium cation is directly added as a salt (DDMA-Br) while for group 2 cations this happens even if the corresponding tertiary amine (DDMA) is added, instead of DDMA-Br. This is attributed to the fact that only group 2 cations can facilitate the protonation of DDMA by the excess oleic acid present in the reaction environment. In all cases studied, the incorporation of M cations is marginal and the reshaping of the NCs is only transient: if the reactions are run for a long time the truncated cubes evolve to cubes.

INTRODUCTION

Colloidal lead halide perovskite nanocrystals (NCs), with general formula APbX$_3$ (A being a large monovalent cation and X being Cl, Br, or I), have been investigated extensively since their first synthesis reports.[1,2] This is due to their excellent optical properties, including high photoluminescence (PL) quantum yield (QY), narrow PL linewidth and tunable PL emission/optical bandgap. The latter can be adjusted by varying the halide and/or A-cation composition and the NCs size.[3-6] To further modulate their optical properties, considerable efforts have also been devoted to controlling the shape of these NCs, resulting in platelets, wires, multipods, and various types of polyhedra.[7-12] In this context, we specifically discuss CsPbBr$_3$ NCs as the main representatives of this field. Colloidal CsPbBr$_3$ NCs are typically synthesized using a hot-injection strategy that involves carboxylic acids (e.g. oleic acid), alkylamines (e.g. oleylamine), high-boiling solvents (e.g. octadecene, hexadecane or diphenyl ether), Cs$^+$ and Pb$^{2+}$ salts, and a source of bromide ions.[1,13,14] Alkylamines, although commonly employed, are not essential for the synthesis, and monodisperse CsPbBr$_3$ cubes can also be prepared using oleic acid alone or in combination with a phosphine oxide, as shown by some of us in previous works.[15,16] In such cases, the NCs surface is exclusively terminated by Cs-oleate species.[17]

A recurrent and more general aspect in the reported synthesis schemes is the use of an over-stoichiometric amount of Pb$^{2+}$, with the Cs:Pb precursors molar ratio often set to 1:2 or even lower.[1,9,17-19] Although a ratio of 1:1 might seem adequate for the growth of CsPbBr$_3$ NCs, it is likely that this condition, in some of the most popular synthesis routes, did not yield satisfactory results, as it was not reported.[20] Yet, using an over-stoichiometric amount of Pb$^{2+}$ should result in a considerable excess of unreacted Pb$^{2+}$ ions remaining in the growth solution after the precipitation of the NCs. In this work, our first step was to employ a previously reported synthesis approach that uses oleic acid as the sole ligand (referred to as the "minimal synthesis").[16] We confirmed that an over-stoichiometric amount of Pb$^{2+}$ (i.e. Cs:Pb of 1:2) is necessary to obtain monodisperse, cubic-shaped NCs that are colloidally stable.

Subsequently, we explored the possibility to replace the excess Pb$^{2+}$ ions in the "minimal synthesis" with different types of cations

(M), which we will name "exogenous". Our goal was to investigate whether we could still obtain monodisperse and colloidally stable CsPbBr$_3$ NCs, thereby guaranteeing the quantitative consumption of Pb$^{2+}$ ions. This approach is particularly appealing if the exogenous cations are less toxic than Pb$^{2+}$.

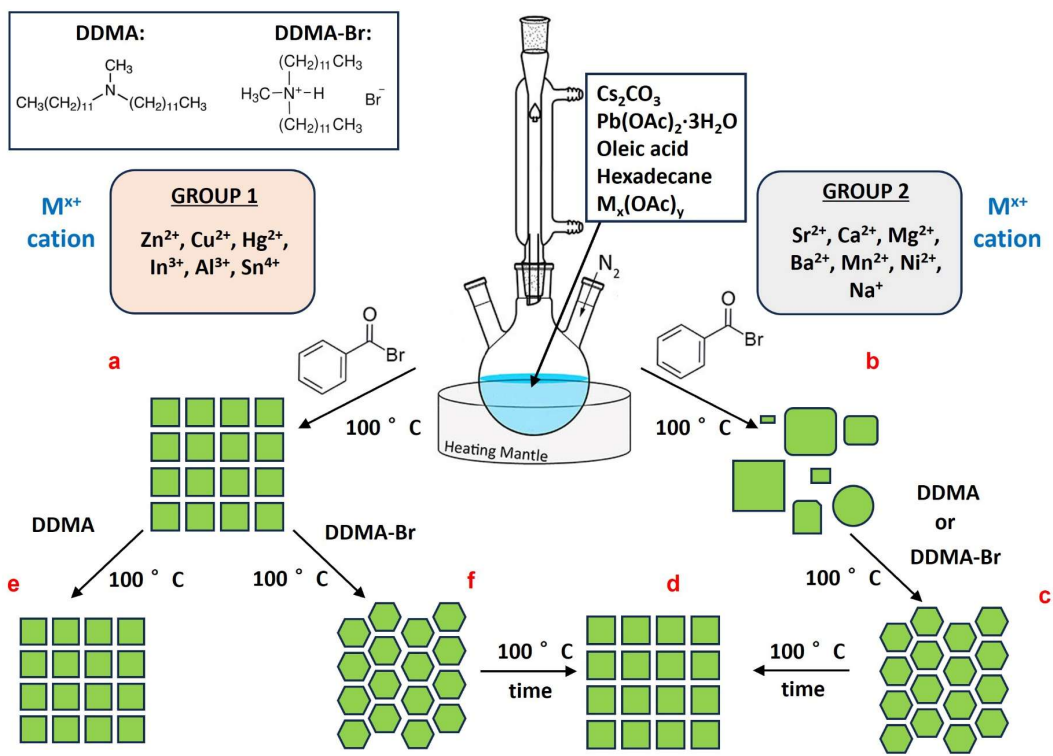

**Scheme 1.** Schematic representation of the "minimal synthesis" approach to CsPbB$_3$ NCs in the presence of exogenous M cations (**a, b**), eventually followed by the addition of either the tertiary amine (didodecylmethyl amine, DDMA) or the corresponding tertiary ammonium bromide (DDMA-Br) (**c, e, f**), and then, for the (**c**) and (**f**) cases, after a prolonged reaction time.

We examined a broad range of M cations (including monovalent, divalent, trivalent and tetravalent ones), which we could categorize into two groups based on the synthesis outcomes. Group 1 comprises Zn$^{2+}$, Hg$^{2+}$, Cu$^{2+}$, In$^{3+}$, Al$^{3+}$ and Sn$^{4+}$. All these cations yielded monodisperse CsPbBr$_3$ nanocubes ranging in size from 8 to 16 nm, depending on the cation used, as shown in **Scheme 1a**. These results are comparable to the "minimal synthesis" scheme based on the 1:2 Cs:Pb precursors ratio run in the absence of exogenous cations. All these cations appear to have in common with Pb$^{2+}$ the ability to form relatively strong complexes with oleates.[21-24] In all the cases, there was a nearly complete consumption of Pb$^{2+}$ ions during the formation of CsPbBr$_3$ NCs, indicating that the synthesis was highly efficient. Indeed, once the NCs were precipitated from the crude reaction solution, the supernatant contained primarily the M-oleate complexes as the unreacted metal species. The incorporation of M cations in the NCs was marginal, typically limited to a few percent, and likely occurring as surface adsorbates/dopants. Notably, some of the group 1 cations resulted in oleate capped NCs with high PLQY values (80-90% for In$^{3+}$, Al$^{3+}$, Zn$^{2+}$), well above those reported for systems with a oleate surface termination.[15-17] It is also important to mention that the PLQY of the NCs synthesized using a Cs:Pb ratio of 1:2, with no exogenous cations, as discussed earlier, was only 40%. A second group of exogenous cations (Sr$^{2+}$, Ni$^{2+}$, Mg$^{2+}$, Ca$^{2+}$, Ba$^{2+}$, Mn$^{2+}$, Na$^+$) yielded NCs that were challenging to purify and precipitate quantitatively from the crude reaction solution. The NCs that we managed to isolate from these reactions exhibited a broad size distribution, as depicted in **Scheme 1b**. A common characteristics of these group 2 cations is their tendency to form weak complexes with oleates.[21-24] A significant advancement of this work was then the discovery that, even with group 2 cations, it was possible to obtain monodisperse NCs featuring high PLQY values (ranging from 70 to 90%, depending on the specific M cation used) by injecting a tertiary amine (namely didodecylmethyl amine, DDMA) in the reaction mixture, immediately after the NCs formation (**Scheme 1c**), rather than a primary or secondary amine. Initially, this led to NCs with truncated cubic shape, but extending the reaction time after amine injection resulted in a transition to cubic shaped NCs, as shown in **Scheme 1d**. This route as well led to nearly complete consumption of Pb$^{2+}$ ions, with negligible incorporation of the exogenous cations in the NCs, and the latter could be easily precipitated and purified at the end of the synthesis.

To explore the role of DDMA further, we extended the same procedure to group 1 cations. In these cases, the addition of the tertiary amine did not alter the shape of the NCs, which remained cubic (**Scheme 1e**). Previous extensive studies have evidenced that neutral alkylamines interact only weakly with the surface of CsPbBr$_3$ NCs, whereas alkyl ammonium ions form much stronger interactions.[17, 25, 26] Based on this, we hypothesized that DDMA undergoes significant protonation only in the presence of group 2 cations, which enables it to reshape the NCs. This hypothesis was reinforced by control experiments in which we deliberately injected tertiary ammonium ions, in the form of didodecylmethyl ammonium bromide (DDMA-Br), into the reaction mixture. When using DDMA-Br, truncated cubes were consistently formed, regardless of the presence of group 1 or 2 cations, as depicted in **Scheme 1c,f**. To better rationalize these findings, we conducted extensive analyses, including nuclear magnetic resonance (NMR) spectroscopy and computational modeling, along with additional control syntheses. These analyses clarified the several key points:

i) The degree of protonation of DDMA by oleic acid, when compared to a simple mixture of the two molecules, decreased in the presence of group 1 cations, but increased with group 2 cations.

ii) Once tertiary ammonium ions are introduced into the reaction (either through direct DDMA-Br addition or by DDMA protonation), they bind to the surface of NCs. In contrast, neutral tertiary amines do not show significantly binding to the NCs.

iii) The rapid increase in tertiary ammonium ions concentration causes etching of the NCs, leading to their initial transformation to truncated cubes. However, this shape change is transient; allowing the synthesis to continue results in smaller NCs with a cubic shape (**Scheme 1c, d, f**) and with a surface enriched in DDMA$^+$ cations.

This work demonstrates that the synthesis of CsPbBr$_3$ NCs can be optimized by replacing the over-stoichiometric amount of Pb$^{2+}$ cations with other exogenous M cations, often resulting in high PLQY values. Additionally, these different cations can influence the protonation of tertiary amines (when introduced to the reaction environment), which in turn can impact the NCs morphology and the composition of the ligand shell.

## RESULTS AND DISCUSSION

Our "minimal synthesis" scheme is an adaptation of a consolidated hot-injection method.[16, 19] We employed oleic acid as the sole ligand, hexadecane as the solvent, benzoyl bromide as the Br precursor, the reaction time was set to 1 min and we varied the Cs:Pb precursors ratio.[19] Details of all syntheses are reported in the Experimental Section.

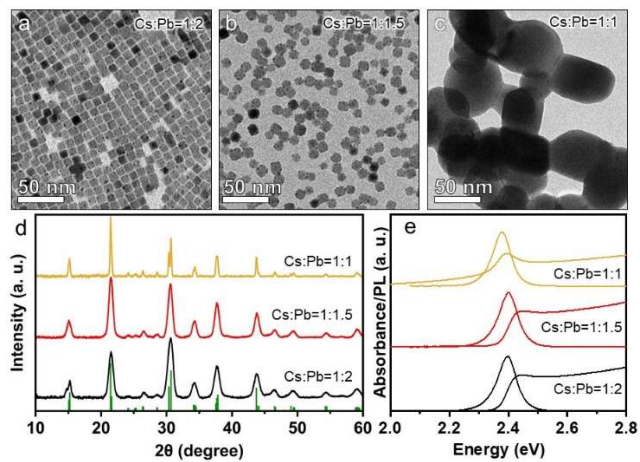

**Figure 1.** (a-c) Transmission electron microscopy (TEM) images, (d) X-ray powder diffraction (XRD) patterns, and (e) optical absorption and PL emission spectra of the NCs prepared at Cs:Pb precursors ratios equal to 1:2, 1:1.5 and 1:1. In (d) the reflections of bulk CsPbBr$_3$ (ICSD number 97851) are represented by vertical green bars.

This synthesis scheme was effective for a Cs:Pb molar ratio of 1:2, delivering colloidally stable, monodisperse cubes with an average size of 12.1±0.6 nm (**Figure 1a**) and a PLQY of 37% (**Figure S1**). The yield of the synthesis, defined in terms of the moles of Pb$^{2+}$ incorporated in the NCs, was assessed using optical absorption spectroscopy. For this, we used the size-dependent molar extinction coefficient of CsPbBr$_3$ NCs reported by Maes et al.[27] The calculated yield was ~46%, indicating that 54% of the initial Pb$^{2+}$ ions remained unreacted in solution (**Figure S2a**). When we used lower amounts of Pb (or, equivalently, higher Cs:Pb ratios) the results were less satisfactory. For instance, at a Cs:Pb ratio of 1:1.5, the resulting NCs lacked well-defined shapes and tended to aggregate (**Figure 1b**). In this case the synthesis yield was ~63% (**Figure S2b**). At a 1:1 ratio, the NCs were significantly agglomerated and overly large (> 50nm, **Figure 1c**), making yield assessment by optical spectroscopy challenging due to strong light scattering of the suspensions. In all experiments, the NCs exhibited an orthorhombic CsPbBr$_3$ crystal structure without secondary phases, as confirmed by X-ray powder diffraction (XRD, **Figure 1d**) measurements. The absorption and PL spectra of the three NC samples showed the typical signatures of CsPbBr$_3$ (**Figure 1e**). These experiments evidenced that controlling the colloidal stability and size distribution of CsPbBr$_3$ NCs is feasible only with a substantial over-stochiometric amount of Pb precursors. This confirms the general synthesis trend observed in the literature.[1, 9, 18, 19]

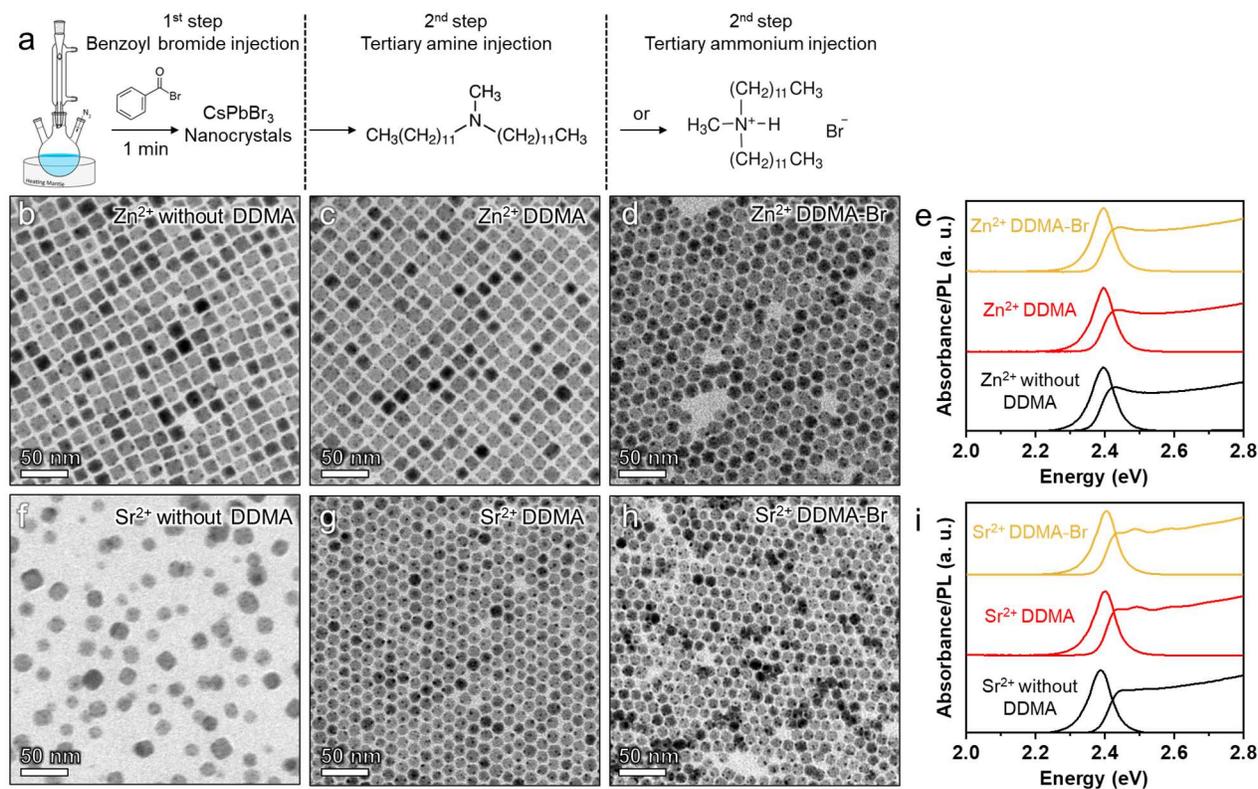

**Figure 2.** (a) Sketch of the synthesis schemes of CsPbBr$_3$ NCs using either Zn$^{2+}$ or Sr$^{2+}$ as exogenous cations, employing oleic acid as the only surfactant (b, f), and after adding either the tertiary amine (c, g) or the corresponding ammonium bromide salt (d, h) once the NCs had formed. For the syntheses of panels b and f, in which neither DDMA nor DDMA-Br was injected, the reaction time was set to 1 min. In all the other cases, the reaction was quenched 10 sec after the DDMA (or DDMA-Br) injection. (e, i) Optical absorption and PL spectra of all the corresponding samples. All the syntheses delivered phase pure CsPbBr$_3$ NCs.

We then explored the possibility to replace the over-stochiometric amount of Pb$^{2+}$ with other exogenous M cations. We categorized these cations into two groups, based on the outcomes of the respective syntheses, as will be clarified later. Group 1 includes Zn$^{2+}$, Hg$^{2+}$, Cu$^{2+}$, In$^{3+}$, Al$^{3+}$, and Sn$^{4+}$, while Group 2 consists of Sr$^{2+}$, Ni$^{2+}$, Mg$^{2+}$, Ca$^{2+}$, Ba$^{2+}$, Mn$^{2+}$, and Na$^{+}$. Our discussion initially focuses on Zn$^{2+}$ and Sr$^{2+}$ cations as representatives of the Group 1 and 2, respectively, with a generalization to the other cations in later sections of this work.

When substituting the over-stochiometric amount of Pb$^{2+}$ with Zn$^{2+}$, specifically working with a Cs:Pb:Zn ratio of 1:1:1, the synthesis yielded uniform cubic shaped CsPbBr$_3$ NCs with a size of 14.4 nm (**Figure 2b**). These NCs exhibited a PLQY of ~80%, which remained stable for at least one month when the NCs were stored under inert atmosphere (**Figure S3**). This is a remarkable result, especially when compared to CsPbBr$_3$ NCs coated with oleates, which usually have lower PLQY values (maximum 60%,[15-17] see also the Cs:Pb = 1:2 case discussed above, featuring a PLQY of only 37%).[28] The yield in terms of Pb was nearly 100% (**Figure S4a**), indicating that this strategy ensures complete consumption of the Pb$^{2+}$ precursor, while leaving in solution only Zn$^{2+}$ complexes.

Under the same synthesis conditions, but this time using Sr$^{2+}$ instead of Zn$^{2+}$, we encountered several challenges. The resulting NCs proved difficult to clean and precipitate due to their tendency to degrade during the process. In their crude reaction solution, these NCs featured a cubic shape with smoothed edges and a broad size distribution (**Figure 2f**). Optical characterizations of the diluted crude NC solutions evidenced the typical absorption and PL spectra of CsPbBr$_3$ (**Figure 2i**). This setback encouraged us to identify strategies to improve the synthesis with Sr$^{2+}$ cations. We tested the effect of injecting a primary, secondary or tertiary amine (and specifically oleylamine, didodecylamine, and DDMA, respectively) into the reaction flask after the NCs had formed (i.e. 1 min after the benzoyl bromide injection), and then quickly quenching the reaction within 10 seconds. The use of oleylamine led to CsPbBr$_3$ NCs with a cubic shape and 9.4 nm size, however with poor optical properties (PLQY of 30%, **Figure S5**). In the case of the secondary amine, we experienced the same problems encountered when working with oleic acid alone: that is, the NCs featured a broad size distribution and a not well-defined shape (**Figure S5**). Interestingly, the use of the tertiary amine DDMA (see **Figure 2a** and the Experimental Section) resulted in NCs with a truncated cubic shape, uniform size distribution (**Figure 2g**) and optimal optical properties (92% PLQY, see **Table 1** and **Figure 2i**). Such shape was also observed when doubling either the amount of DDMA injected or the amount of oleic acid (**Figure S6**). The addition of DDMA had therefore induced a digestive ripening of the NCs. The remarkable impact, in terms of unusual NC shape and efficient PL emission, obtained with DDMA in our synthesis scheme, compared to that of either primary or secondary amines, motivated us to further investigate its effects. We therefore tested whether the addition of

DDMA would have a similar impact in the presence of other exogenous cations. However, when we added DDMA to the $Zn^{2+}$-based synthesis, there was no shape change of the NCs, which remained cubic (**Figure 2c** and **Table 1**).

We conducted NMR analyses on the products of the syntheses, specifically those using $Sr^{2+}$ and $Zn^{2+}$ that included DDMA addition, to gain insights into the amine's role. The detailed analyses are discussed in the SI and the results are shown in **Figures S7-17**. The NMR data revealed that the NCs synthesized with $Zn^{2+}$ were only coated by oleates (**Figures S9-10** and **S15** and **Table 1**). In contrast, the NCs synthesized with $Sr^{2+}$ had both oleates and $DDMA^+$ ions bound to their surface (**Figure 3** and **Figures S7, 10-12**): i) the $^1H$–$^{13}C$ heteronuclear single quantum coherence (HSQC) spectrum of Sr-based NCs revealed that the $CH_2$-N proton peaks of DDMA at 2.34 and 2.13 ppm (with the same $^{13}C$ at 54.79 ppm) (**Figure 3a**) were not magnetically equivalent, thus consistent with DDMA being protonated; ii) the $^1H$ nuclear Overhauser effect spectroscopy (NOESY) spectrum of Sr-based NCs evidenced that both $DDMA^+$ ($^1H$ peaks in the 2.1-2.4ppm range, **Figure 3b**) and oleate ($^1H$ peaks at ~5.5 ppm, **Figure 3b**) returned negative (red) NOE cross peaks, which are indicative of species with a slow tumbling regime in solution (i.e. they are bound to the NCs' surface). The quantitative NMR analysis, performed after dissolving the NCs in DMSO (**Figure S10**),[29] returned a $DDMA^+$/oleate ratio of approximately 1.77 (see **Table 1**). Notably, the $CH_2N$ $^1H$ peaks of $DDMA^+$ in DMSO were not only shifted to a lower field compared to those of neutral DDMA but were also split in two multiplets at 2.34 and 2.13 ppm, with the same $^{13}C$ at 54.79 ppm (**Figures S13-15** and SI for further details). Considering that the $CH_2N$ protons of neutral DDMA in DMSO exhibited a single broad peak at high field with the $^1H$ resonance at 2.09 and the $^{13}C$ at 57.08 ppm (**Figure S13**), our NMR analyses indicate that the $CH_2N$ protons in $DDMA^+$ are not magnetically equivalent. We tentatively ascribe the non-equivalence of the $CH_2N$ protons in $DDMA^+$ to both the steric hindrance on the nitrogen atom (due to the methyl and didodecyl substituents) and to the protonation, which hinders the inversion process on nitrogen (possible only in the presence of an electron lone pair).[30]

These findings suggest that a significant protonation of DDMA, resulting in $DDMA^+$ cations, occurred exclusively in the presence of $Sr^{2+}$, and not for $Zn^{2+}$. To support this hypothesis, we conducted additional NMR analyses on a mixture of DDMA and oleic acid, as well as on the same mixture in the presence of either $Zn^{2+}$ or $Sr^{2+}$, at various temperatures (25°C, 50°C and 80°C, see **Figure S17**). The results consistently showed that the protonation of DDMA was enhanced by $Sr^{2+}$, while $Zn^{2+}$ inhibited it, across all temperatures tested. We discuss a possible explanation of this effect later in this work.

**Table 1. NC size, PLQY and $DDMA^+$/oleate ratio in the ligand shell for syntheses carried out under the various reaction conditions.**

| Molecule injected and quenching time | Shape | Size [nm] | PLQY [%] | $\frac{DDMA^+}{Oleate}$ ratio |
|---|---|---|---|---|
| $Sr^{2+}$ case | | | | |
| DDMA 10 sec | Truncated Cubic | 12.1 ± 0.5 | 92 | 1.77 |
| DDMA-Br 10 sec | Truncated Cubic | 10.2 ± 0.5 | 97 | 1.75 |
| DDMA 20 min | Cubic | 8.1 ± 0.5 | 84 | 5.14 |
| DDMA-Br 20 min | Cubic | 7.1 ± 0.4 | 91 | 2.63 |
| $Zn^{2+}$ case | | | | |
| DDMA 10 sec | Cubic | 13.8 ± 0.5 | 81 | no $DDMA^+$ |
| DDMA-Br 10 sec | Truncated Cubic | 12.7 ± 0.5 | 86 | 1.71 |
| DDMA-Br 20min | Cubic | 9.0 ± 0.5 | 76 | 2.05 |

In the $Sr^{2+}$ case, the $DDMA^+$ cations formed are believed to be responsible for the etching of the NCs and for the partial displacement of oleate ligands from their surface. This observation led us to hypothesize that the direct injection of $DDMA^+$ cations into the reaction environment would induce a reshaping of the NCs regardless of the type of M cations present. Following this line of reasoning, the syntheses with $Sr^{2+}$ and $Zn^{2+}$ cations were repeated, and this time we injected a tertiary ammonium bromide salt (i.e. DDMA-Br dissolved in hexadecane) instead of the neutral DDMA. In both cases, the resulting NCs exhibited a truncated cubic shape (**Figure 2d, h**) and were passivated by a mixture of oleates and $DDMA^+$ ions, with the $DDMA^+$/oleate ratio being around 1.7 (**Table 1** and **Figures S18-19**). These NCs exhibited high PLQY values (>85%, see **Figure S20**). These findings confirmed our initial hypothesis that a significant protonation of DDMA, leading to the formation of $DDMA^+$ cations, is crucial to induce the observed shape change.

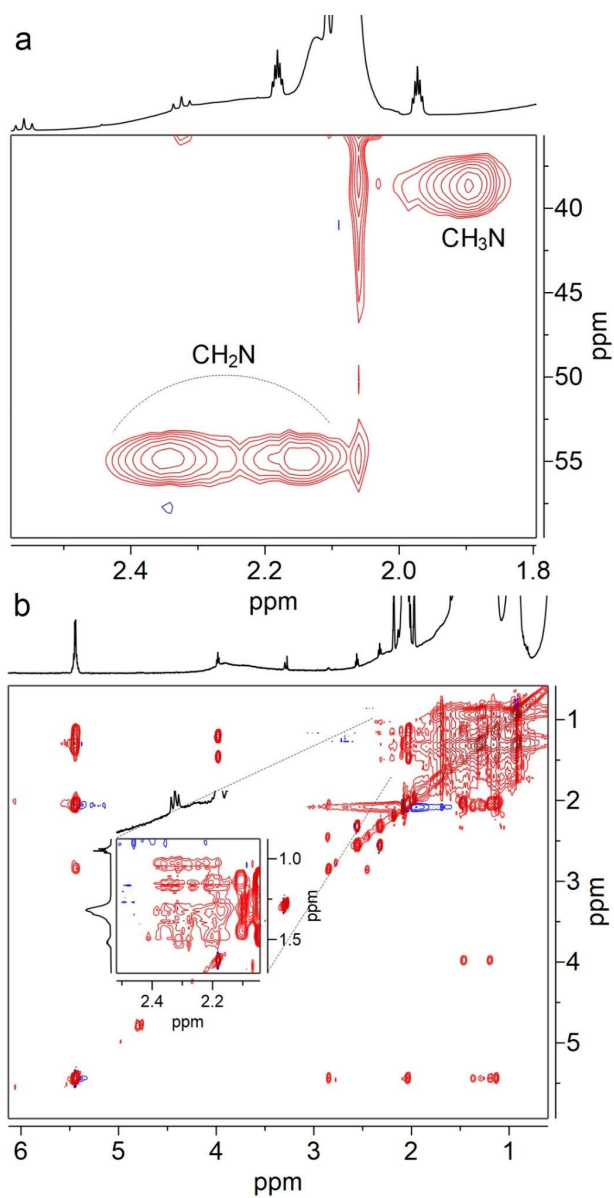

**Figure 3.** a) $^1$H–$^{13}$C HSQC and b) the $^1$H-$^1$H NOESY spectra of colloidal suspensions of CsPbBr$_3$ NCs synthesized in the presence of Sr$^{2+}$ (with the injection of the DDMA, reaction quenched after 10 sec) in toluene-D.

In all the experiments discussed so far, the products were confirmed to be CsPbBr$_3$ NCs, as assessed by optical (**Figure 2e,i**) and XRD (**Figure S21**) analyses, regardless of the varied reaction conditions. Additionally, elemental analysis using energy dispersive X-ray spectroscopy (EDS), performed in a scanning electron microscope (SEM), revealed no significant incorporation of the exogenous cations in the NCs (**Figure S22**). It is also worth noting, as previously mentioned, that a consistent and important finding across all these syntheses was the nearly complete incorporation of Pb$^{2+}$ ions in the final CsPbBr$_3$ NCs (**Figures S4** and **S23**).

To gain further insights into the etching and reshaping process, we conducted high-angle annular dark field-scanning TEM (HAADF-STEM) analysis on a selected sample. Specifically, we examined the truncated cubes synthesized with Sr$^{2+}$ and DDMA, as shown in **Figure 4**. The STEM image (**Figure 4a**) is compatible with the [001] projection of a NCs having either {111} or {110} type facets truncations (as illustrated in **Figure 4b**), assuming a pseudo-cubic cell for simplicity in describing the perovskite structure, despite the orthorhombic cell being a better fit for the experimental data. STEM-EDS maps (**Figure 4c** and **Figure S24**) revealed that the Sr signal (corresponding to only a 2.5% atomic percentage, consistent with the SEM-EDS results) is not colocalized with the NCs. This suggests that this detected Sr is likely due to residual, unwashed Sr-oleate rather than its incorporation into the NCs in meaningful amounts.

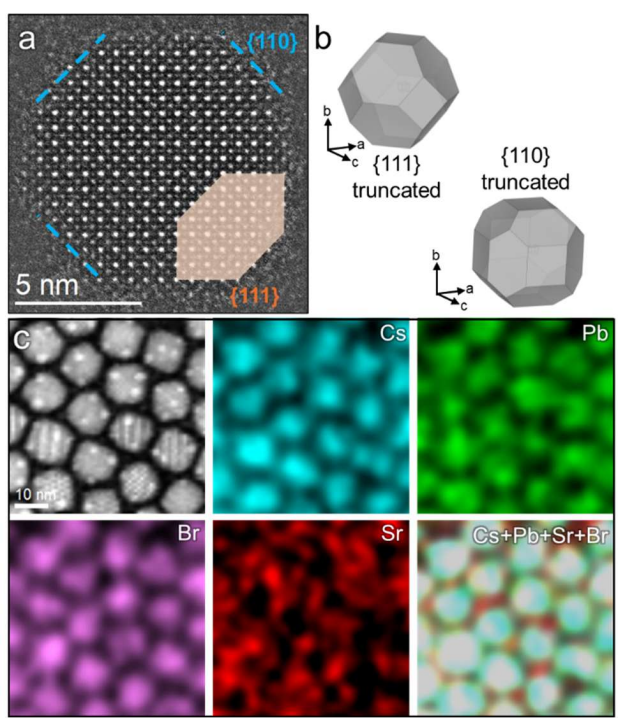

**Figure 4.** (a) High-angle annular dark field-scanning TEM (HAADF-STEM) and (c) STEM-EDS analysis of the truncated cube samples corresponding to the synthesis of Figure 2g. (b) models of the truncated cubes with either (111) or (110) facets truncations.

To further explore the reshaping process, we repeated the syntheses involving Sr$^{2+}$+DDMA, Sr$^{2+}$+DDMA-Br, and Zn$^{2+}$+DDMA-Br, allowing them to proceed for 20 minutes instead of quenching them after 10 seconds. What we discovered is that the etching/reshaping of NCs into truncated octahedra is transient, and leads, eventually, to smaller cubic-shaped NCs (**Figure 5**). This size reduction and shape transition (see **Table 1**, **Figure 2** and **5**) further indicated that DDMA$^+$ ions played a crucial role in the etching process. This conclusion is supported by the NMR analyses, which showed that the final cubic NCs, compared to the earlier truncated ones, had a surface termination that was further enriched in DDMA$^+$ (**Table 1** and **Figures S25-27**). Again, in all these syntheses the reaction yield for Pb incorporation in the NCs was close to 100% (see **Figure S28**).

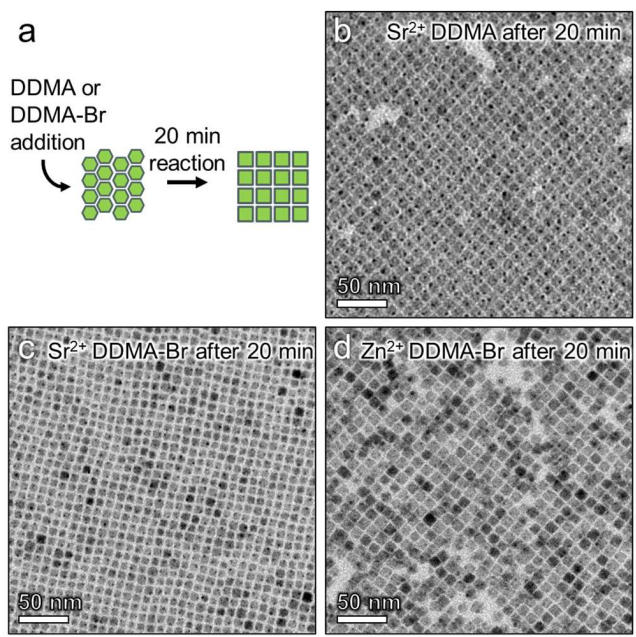

**Figure 5.** TEM images of the NCs synthesized using either $Zn^{2+}$ or $Sr^{2+}$ as exogenous cations, after injecting either (b) DDMA (for the $Sr^{2+}$ case) or (c, d) DDMA-Br (for the $Sr^{2+}$ and $Zn^{2+}$ cases) and then letting the reaction run for 20 min. In all the syntheses, the DDMA/DDMA-Br addition initially produced truncated cubes. However, after prolonged reaction time the shape evolved to cubes (a).

In an attempt to rationalize the etching process and the presence of DDMA$^+$ ions on the surface of truncated cubic shaped NCs, we performed ligand binding affinity calculations at the density functional theory (DFT) level. We emphasize that the binding affinity of primary and quaternary (specifically didodecyldimethyl ammonium) alkylammonium cations toward CsPbBr$_3$ NCs is well-known and thoroughly documented in the literature.[28, 31-33] Conversely, it has been shown experimentally that secondary alkylammonium cations such as protonated dihexylamine and dioctadecylamine present an unfavorable binding to these NCs, as reported by Imran et al.[16] Yet, no studies have been made on tertiary alkylammonium ions, such as DDMA$^+$ discussed in the present work. Here, we considered a simplified version of DDMA$^+$ ligands, featuring shorter alkyl chain such as diethyl methyl ammonium and a 2.4 nm-sided cubic CsPbBr$_3$ NC model employed previously in other works.[28, 34, 35] The binding affinity of the ligands to the NC surface was defined as:

$$\Delta E_{binding} = E_{Complex} - (E_{Core} + E_{Ligand}) \quad (1)$$

where *Complex* is the NC core passivated with one ligand, *Core* is the NC core with a *Ligand* vacancy and *Ligand* is the dissociated ligand. In our simulations, *Ligand* is either an original CsBr ion pair (that mimics a more generic CsX pair with X=Br, oleate), a neutral tertiary amine or a tertiary ammonium-Br (TABr) ion pair. The calculated values are reported in **Table 2** and the associated binding sites are depicted in **Figure S29**. The neutral tertiary amine features a binding energy of only 2.14 kcal/mol, indicating the weak interaction of this species with the surface and presumably its inability to cause any shape change of the NC. On the other hand, our DFT calculations revealed an exothermic binding energy value of 33.54 kcal/mol for TABr, which is similar to those calculated for primary and secondary ammonium ions (**Table 2**). This suggests that alkylammonium ions may indeed efficiently passivate CsPbBr$_3$ NCs, and eventually modulate their shape.

**Table 2. Ligands binding affinity values on the 001 facets of a 2.4 nm-sided cubic CsPbBr$_3$ NC model, from DFT calculations.**

| Ligand | $\Delta E_{binding}$ (kcal/mol) |
| --- | --- |
| CsBr | -52.27 |
| Tertiary Amine (neutral) | -2.14 |
| Tertiary Ammonium-Br (TABr) | -33.54 |
| Secondary Ammonium-Br | -36.15 |
| Primary Ammonium-Br | -42.89 |

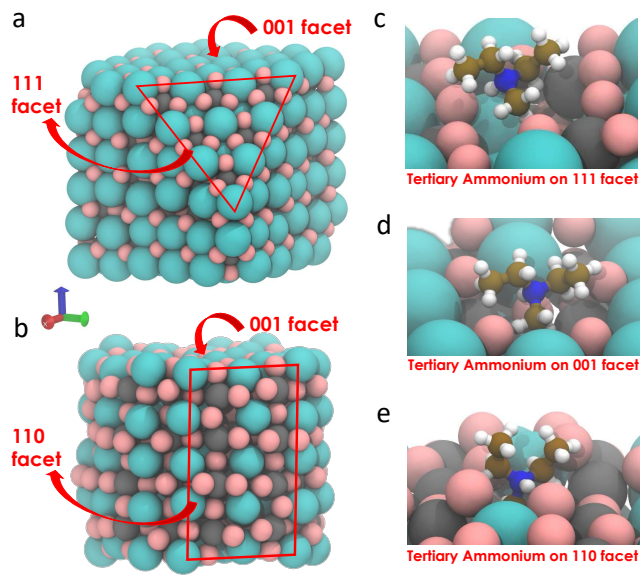

**Figure 6.** (a, b) Ball representation of a 2.4nm CsPbBr$_3$ NC model exposing, respectively, six (001) facets as well as two (111) facets, and two (001) facets as well as six (110) facets in a cubic lattice representation. (c, d, e) Sketches of the binding features for a tertiary ammonium ligand anchored, respectively, at the (111), (001) and (110) facet types. The following color code is employed: Cs: cyan; Pb: grey; Br: pink; C: brown; N: blue; H: white.

We also developed two truncated NC models exposing, respectively, both the (001) and (111) facets (**Figure 6a**) and both the (001) and (110) facets (**Figure 6b**), in line with the experiments. The binding of TABr on the (111) facet is only ~2 kcal/mol more favorable than on the (001) facets (binding sites represented in **Figure 6c, d**). Similarly, TABr stabilizes the (001) and (110) facets almost equally in terms of energy, with the computed energy difference being only ~0.1 kcal/mol (binding sites are represented in **Figure 6d, e**). Moreover, we calculated that the binding of the CsX ion pair at the NC surface is energetically favored over its TABr counterpart (by ca. 20 kcal/mol). This energy difference in favor of CsX pairs may indicate that the binding of TA cations could involve a two-step mechanism. We speculate that this process might start with an etching of the outer layers, presumably

CsX/PbBr$_2$/CsBr layers, caused by TA cations, which then leads to their subsequent binding to the NC surface.

It is important to notice that these calculations only account for the effect of one anchoring group on the NC surface, while neglecting the ligand-to-ligand and ligand-to-solvent interactions, both of which are known to strongly affect the free binding energy of ammonium ligands.[16] Furthermore, entropic and kinetic terms are also neglected and may play an important role in differentiating between primary, secondary and ternary ammonium ions. Based on these considerations, we conclude that our calculations cannot explain why tertiary ammonium ions lead to truncated cubes a few seconds after their injection, while primary ammonium ions lead to cubes. We speculate that this could be attributed to kinetic factors. Indeed, for tertiary ammonium ions, the shape obtained by prolonging the reaction time is also cubic. Future studies might shed more light on this puzzle.

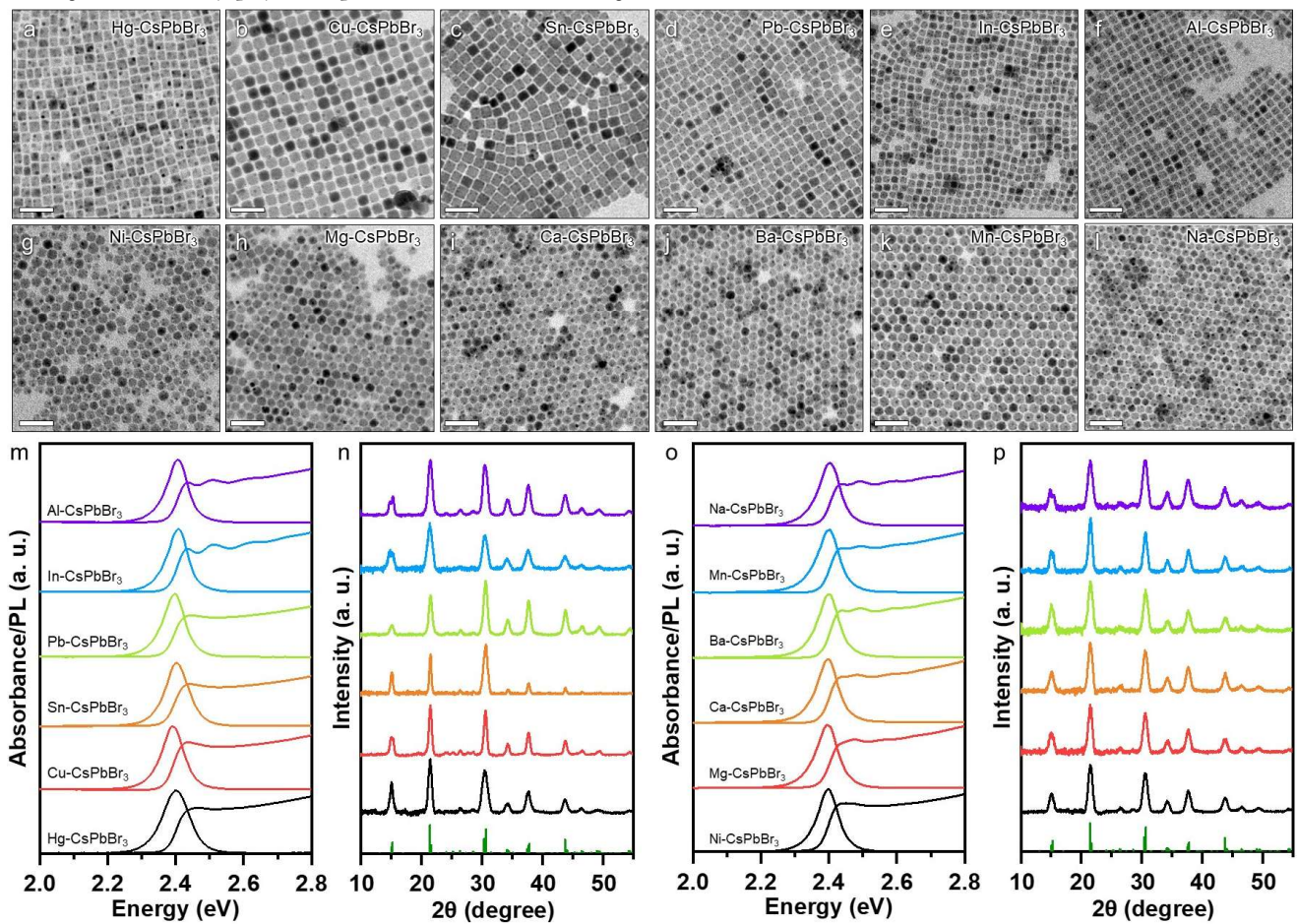

**Figure 7.** (a-l) TEM images of NCs synthesized using various exogenous cations, collected soon after the ternary amine addition. The corresponding (m, o) optical absorption and (n, p) PL characteristics and XRD patterns are reported in panels m-p. The scale bar in (a-l) is 50 nm.

Overall, the experiments discussed above offer a comprehensive picture for the Sr$^{2+}$ and Zn$^{2+}$ cases, but these two cations are not alone in their behavior. We have extended our synthesis approach, which involves the use of an exogenous M cation followed by the addition of DDMA and quenching of the reaction after 10 sec, to a broad range of exogenous cations, including M$^+$, M$^{2+}$, M$^{3+}$ and M$^{4+}$ cations. The results are reported in **Figure 7**. Based on the results, we can categorize these additional cations into two groups. Group 1 includes cations such as Hg$^{2+}$, Cu$^{2+}$, Pb$^{2+}$, In$^{3+}$, Al$^{3+}$, and Sn$^{4+}$, which yielded NCs with a cubic shape after the addition of DDMA (**Figure 7a-f**). Similarly to the Zn$^{2+}$ case, group 1 cations yield nanocubes when employing only oleic acid and truncated nanocubes upon the addition of DDMA-Br (**Figure S30**). Group 2 comprises cations like Ni$^{2+}$, Mg$^{2+}$, Ca$^{2+}$, Ba$^{2+}$, Mn$^{2+}$, Na$^+$, which upon DDMA addition led to NCs with a truncated cubic shape (**Figure 7g-l**, the corresponding NCs sizes are reported in **Table S1**) passivated by both DDMA$^+$ and oleate ions (**Figures S31-32**). Analogously to the Sr$^{2+}$ case, group 2 cations yielded NCs with a poorly defined shape and a broad size distribution when no DDMA (or DMMA-Br) was used (**Figure S30**). Notably, Pb$^{2+}$ is included in Group 1 cations because the synthesis conducted with a Cs:Pb ratio of 1:(1+1) = 1:2 preserves the cubic shape post-amine addition.

One important remark from these experiments is that all the syntheses consistently delivered phase pure CsPbBr$_3$ NCs. This is confirmed by optical absorption and PL spectra, as well as by XRD patterns (**Figure 7m-p** and **Figure S30**). Indeed, the incorporation of exogenous M cations in the resulting CsPbBr$_3$ NCs was minimal, as indicated by the M/Pb ratios measured by ICP-OES, which fell within the range of 0.01 to 0.10 (see Table S1). Alloying of the NCs with a few percent of exogenous cations, or their surface termination by such cations, is a possibility we cannot rule out. This is suggested by the wide range of PLQY values observed, which vary from 1% to 85%, across different samples (**Figures S33-34**). Specifically, the cases with $Hg^{2+}$ and $Cu^{2+}$ exhibit very weak PLQY (~1%), whereas samples with $Zn^{2+}$ and $Al^{3+}$ demonstrate significantly higher PLQY, reaching values as high as 85%. Despite these variations, the consistent behavior observed in terms of shape preservation or change across each series hints at an underlying commonality, that appears to be independent of the specific cation size, charge or electronic configuration.

As anticipated in earlier sections of this work, the primary distinction between Group 1 and Group 2 cations lies in their binding strength to oleates. Previous studies investigating the association tendencies of divalent metal cations with various carboxylate ions identified a general $Zn^{2+} \gg Ca^{2+} > Sr^{2+} > Ba^{2+}$ trend, with the $Mn^{2+}$ position in this trend being somewhat erratic.[22] Bunting and Thong conducted experiments to measure the stability constants for 1:1 complexes formed between various divalent cations and different carboxylate anions. They observed a trend in stability that follows the order $Pb^{2+} > Cu^{2+} > Cd^{2+} > Zn^{2+} > Ni^{2+} \sim Co^{2+} > Ca^{2+} \sim Mg^{2+}$, which aligns with our two-group classification.[36] Robertson and colleagues used a spectroscopic analysis to study the interaction between metal ions and decanoate at an oil-water interface. Their finding indicated that $Mg^{2+}$, $Ca^{2+}$, $Mn^{2+}$ and $Ni^{2+}$ (all group 2 cations in our study) exhibit a purely ionic interaction, whereas $Zn^{2+}$ and $Cu^{2+}$ (group 1 cations in our study) show stronger, more covalent interactions.[37] Furthermore, a recent Raman spectroscopic study highlighted that the binding of $Zn^{2+}$ to acetate is significantly stronger than that of either $Mg^{2+}$ or $Ca^{2+}$.[38] The bonding of carboxylates with alkali metal cations is so weak that mass spectrometry methods are typically employed to study them.

With this in mind, we performed DFT calculation to compute the binding energies between the various cations studied in this work and oleate ions, the latter modelled using shorter alkyl chain ligands, i.e. acetate ions. The binding energy values per carboxylate are listed in ascending order in **Table 3**. The structures of the corresponding M(acetate)$_x$ molecular complexes are reported in **Figure S35**. The energetics associated to the $Ni^{2+}$ and $Mn^{2+}$ cations, presenting an open-shell ground state electronic configuration, were excluded from the analysis. The computed metal-carboxylate binding strengths are closely related to the classification of the cations based on their effect in the reaction mixture. Group 1 cations (those for which NCs tend to preserve a cubic shape and tertiary amines are not protonated in solution, highlighted in magenta in **Table 3**), present higher affinities toward carboxylate ions. Group 2 cations (those for which the NCs are reshaped to truncated cubes upon amine addition, due to the protonation of the latter, highlighted in cyan in **Table 3**) consistently show lower affinities toward carboxylate ions. $Mg^{2+}$ represents the only exception to this trend.

**Table 3. Binding energy values between the various M cations and the oleate ions from the DFT calculation. Group 1 cations are shaded in magenta, Group 2 in cyan.**

| M | Shannon radius (Ang) | $\Delta E_{Binding}$ (Kcal/mol) per carboxylate |
|---|---|---|
| $Sn^{4+}$ | 0.69 | -715.6 |
| $Al^{3+}$ | 0.54 | -453.0 |
| $In^{3+}$ | 0.80 | -421.5 |
| $Cu^{2+}$ | 0.73 | -348.1 |
| $Zn^{2+}$ | 0.74 | -331.3 |
| $Hg^{2+}$ | 1.02 | -317.8 |
| $Mg^{2+}$ | 0.72 | -294.5 |
| $Pb^{2+}$ | 1.19 | -263.8 |
| $Ca^{2+}$ | 1.00 | -257.2 |
| $Sr^{2+}$ | 1.18 | -238.7 |
| $Ba^{2+}$ | 1.35 | -221.9 |
| $Na^+$ | 1.02 | -149.3 |

We now come to the point of trying to provide a rationale for why different cations can influence the protonation of DDMA. Tertiary alkyl amines are easily protonated by alkyl carboxylic acids in an aqueous environment,[39, 40] yet in a low polarity, non-aqueous environment protonation is more difficult. Protonation might be then promoted by specific species. This appears to be the role of complexes involving group 2 cations with oleates, which, by virtue of the weak association constants, might provide active sites for stabilizing the final products of the protonation reaction. Evidently, for the comparatively stronger complexes involving group 1 cations with oleates this role is precluded.

CONCLUSIONS

In this work we have demonstrated a synthesis scheme to CsPbBr$_3$ NCs, wherein the commonly employed over-stoichiometric amount of $Pb^{2+}$ precursor is replaced with exogenous metal cations. The use of these exogenous metal cations enabled the formation of CsPbBr$_3$ NCs with complete consumption of the Pb precursor. Furthermore, our experiments revealed that cations forming strong complexes with oleates could produce cubic NCs with a photoluminescence quantum yield (PLQY) that in some cases could be as high as 90% (i.e. $Al^{3+}$ cations), passivated solely by oleates. Conversely, cations forming weak complexes with oleates could not yield a good control over the size and colloidal stability of the NCs unless a tertiary amine (and not a primary or secondary amine), specifically DDMA, was introduced in the reaction environment. In this last case, we obtained NCs with a truncated cubic shape, a PLQY as high as 90% (i.e. $Ca^{2+}$ and $Na^+$) and a surface passivated by both DDMA$^+$ and oleate species. The same shape was obtained, irrespective of which exogenous cations were used, if instead the ammonium cation was directly added as a salt (DDMA-Br). The shape change was always transient and the NCs evolved to cubes over time, however DDMA$^+$ cations continued to passivate their surface. This work suggests that the protonation of alkylamines can be used to regulate the shape and surface termination (thus also surface traps passivation) of the NCs and can be in principle extended to other systems.

EXPERIMENTAL SECTION

**Chemicals.**

Hexadecane (99%), oleic acid (OA, technical grad, 90%), cesium(I) carbonate ($Cs_2CO_3$, 98%), lead (II) acetate trihydrate ($Pb(OAc)_2 \cdot 3H_2O$, 99.99%), strontium (II) acetate ($Sr(OAc)_2$, 99%), magnesium (II) acetate tetrahydrate ($Mg(OAc)_2 \cdot 4H_2O$, 99.99%), calcium (II) acetate hydrate ($Ca(OAc)_2 \cdot H_2O$, 99%), barium (II) acetate ($Ba(OAc)_2$, 99%), sodium acetate (NaOAc, 99%), manganese (II) acetate tetrahydrate ($Mn(OAc)_2 \cdot 4H_2O$, 99.99%), nickel (II) acetate tetrahydrate ($Ni(OAc)_2 \cdot 4H_2O$, 99.995%), copper (II) acetate hydrate ($Cu(OAc)_2 \cdot H_2O$, 99.99%), zinc (II) acetate dihydrate ($Zn(OAc)_2 \cdot 2H_2O$, 99%), aluminum (III) acetate basic ($Al(OAc)_2(OH)$, 99%), indium (III) acetate ($In(OAc)_3$, 99.99%), Tin(IV) acetate ($Sn(OAc)_4$, 99%), toluene (anhydrous, 99.8%) and toluene-$d_8$ (anhydrous, 99.6%), dimethyl sulfoxide-$d_6$ (DMSO-D, 99.8%), ethanol (99.8%), hydrobromic acid (99%), hydrobromic acid (HBr, 99,99%), trifluoro acetic acid (TFA, 98%) and oleylamine (98%) were purchased from Sigma-Aldrich. Benzoyl bromide (98%), didodecylmethyl amine (DDMA, >85%) and didodecylamine (97%) were purchased from Tokyo Chemical Industry (TCI). All reagents were used as received without any further experimental purification.

**Synthesis of CsPbBr$_3$ nanocrystals (NCs) with the "minimal synthesis" approach.**

*1) Synthesis of CsPbBr$_3$ NCs varying the Cs:Pb precursors ratio (from 1:1 to 1:2)*: 0.05 mmol of $Cs_2(CO_3)$, x mmol (x=0.1, 0.15, 0.2) of $Pb(OAc)_2 \cdot 3H_2O$ were dissolved in 1.5 ml (4.73 mmol) of oleic acid and 6 ml of hexadecane in a 25 ml three-necked flask. The resulting mixture was pumped to vacuum at room temperature for 30 min and at 100 °C for 30 min. The mixture was subsequently placed under a nitrogen atmosphere, and the temperature was raised to 180 °C for 10 minutes to achieve a transparent, clear solution. The solution was then cooled down to 100 °C and a benzoyl bromide solution (obtained by mixing 50 μL of benzoyl bromide (0.42 mmol) in 500 ml of hexadecane) was swiftly injected, triggering the nucleation and growth of the CsPbBr$_3$ NCs. The reaction was quenched after 1 minute by rapidly cooling it down to room temperature using an ice-water bath. The NCs were precipitated by the addition of 12 ml a mixture of methyl acetate and toluene (volume ratio of 2:1) and centrifugation at 6000 rpm for 10 min. The precipitate was redispersed in 2 ml anhydrous toluene and the resulting dispersion was centrifuged at 4000 rpm for 5 min. The supernatant was stored in a nitrogen filled glovebox for further characterizations.

*2) Synthesis of Sr and Zn-CsPbBr$_3$ NCs without DDMA*: 0.05 mmol of $Cs_2(CO_3)$, 0.1 mmol of $Sr(OAc)_2$ or $Zn(OAc)_2$, and 0.1 mmol of $Pb(OAc)_2 \cdot 3H_2O$ were dissolved in 1.5 ml (4.73 mmol) of oleic acid and 6 ml of hexadecane in a 25 ml three-necked flask. The resulting mixture was pumped to vacuum at room temperature for 30 min and at 100 °C for 30 min. The mixture was subsequently placed under a nitrogen atmosphere, and the temperature was raised to 180 °C for 10 minutes to achieve a transparent, clear solution. The solution was then cooled down to 100 °C and a benzoyl bromide solution (obtained by mixing 50 μL of benzoyl bromide (0.42 mmol) in 500 ml of hexadecane) was swiftly injected, triggering the nucleation and growth of the CsPbBr$_3$ NCs. The reaction was quenched after 1 minute by rapidly cooling it down to room temperature using an ice-water bath. Zn-CsPbBr$_3$ NCs were precipitated by the addition of 12 ml of a mixture of methyl acetate and toluene (volume ratio of 2:1) and centrifugation at 6000 rpm for 10 min. The precipitate was redispersed in 2 ml anhydrous toluene and the resulting dispersion was centrifuged at 4000 rpm for 5 min. The supernatant was stored in a nitrogen filled glovebox for further characterizations. The same washing procedure led to the degradation of Sr-based NCs. To acquire TEM pictures of these NCs we added anhydrous toluene to the crude reaction solution (volume ratio of crude reaction mixture:toluene was 1:3) and proceeded with the TEM grid preparation.

*3) Synthesis of M-CsPbBr$_3$ NCs with DDMA*: 0.05 mmol of $Cs_2(CO_3)$, 0.1 mmol of $Pb(OAc)_2 \cdot 3H_2O$ and the desired amount of M exogenous cations (in the form of $M_x(OAc)_y$) were dissolved in 1.5 ml (4.73 mmol) of oleic acid and 6 ml of hexadecane in a 25 ml three-neck flask. The amount of $M_x(OAc)_y$ was 0.1 mmol for $M^{2+}$ cations, 0.2 mmol for $M^+$ cations and 0.067 mmol for $M^{3+}$ cations (thus charge balancing the over stoichiometric amount of $Pb^{2+}$ not employed in these syntheses, that is 0.1 mmol of $Pb^{2+}$). The resulting mixture was pumped to vacuum at room temperature for 30 min and at 100 °C for 30 min. The mixture was subsequently placed under a nitrogen atmosphere, and temperature was raised to180 °C for 10 minutes to achieve a transparent, clean solution. The solution was then cooled down to 100 °C and a benzoyl bromide solution (obtained by mixing 50 μL of benzoyl bromide (0.42 mmol) in 500 ml of hexadecane) was swiftly injected, triggering the nucleation and growth of CsPbBr$_3$ NCs. After 1min of reaction, a tertiary amine solution (obtained by mixing 100 μL DDMA (0.2 mmol) dispersed in 900 μL hexadecane) was swiftly injected, and the reaction was quenched within 10 sec by rapidly cooling it down to room temperature using an ice-water bath. The NCs were precipitated by the addition of 12 mL a mixture of methyl acetate and toluene (volume ratio of 2:1) and centrifugation at 6000 rpm for 10 min. The precipitate was redispersed in 2 ml anhydrous toluene and the resulting dispersion was centrifuged at 4000 rpm for 5 min. The supernatant was stored in a nitrogen filled glovebox for further characterizations. The syntheses of Sr-CsPbBr$_3$ NCs with oleylamine and didodecylamine were conducted using the same protocol (that is replacing the 0.2 mmol of DDMA with either oleylamine or didodecylamine). Since didodecylamine is a solid, it was dissolved in hexadecane and heated to 100 °C to form a transparent solution.

*4) Synthesis of Sr- and Zn-CsPbBr$_3$ NCs with DDMA-Br*: 0.05 mmol of $Cs_2(CO_3)$, 0.1 mmol of $Sr(OAc)_2$ or $Zn(OAc)_2$, and 0.1 mmol of $Pb(OAc)_2 \cdot 3H_2O$ are dissolved in 1.5 ml (4.73 mmol) of oleic acid and 6 ml of hexadecane in a 25 ml three-neck flask. The resulting mixture was pumped to vacuum at room temperature for 30 min and at 100 °C for 30 min. The mixture was subsequently placed under a nitrogen atmosphere, and the temperature was raised to 180 °C for 10 minutes to achieve a transparent, clean solution. The solution was then cooled down to 100 °C and a benzoyl bromide solution (obtained by mixing 50 μL of benzoyl bromide (0.42 mmol) in 500 ml of hexadecane) was swiftly injected, triggering the nucleation and growth of the $CsPbBr_3$ NCs. After the reaction was run for 1 min, a tertiary ammonium bromide solution (obtained by dissolving 90 mg *DDMA-Br* (0.2 mmol) in 900 μL toluene) was swiftly injected in the flask, and the reaction was quenched within 10 sec by rapidly cooling it down to room temperature using an ice-water bath. The NCs were precipitated by the addition of 12 mL a mixture of methyl acetate and toluene (volume ratio of 2:1) and centrifugation at 6000 rpm for 10 min. The precipitate was redispersed in 2 ml anhydrous toluene and the resulting dispersion was centrifuged at 4000 rpm for 5 min. The supernatant was stored in a nitrogen filled glovebox for further characterizations.

*5) Synthesis of the didodecylmethyl ammonium bromide (DDMA-Br) salt*: The preparation of ammonium salt was followed from procedures from literature.[39] 2.5 ml DDMA (5 mmol) was added to 10 ml ethanol in a single neck round bottom flask immersed in an ice-water bath. 5 mmol hydrobromic acid was then added dropwise and the reaction was allowed to proceed for 1 hour. The solvent was then removed from the precipitate using a rotary evaporator, and the solids were washed three times with diethyl ether. A white powder was recrystallized in diethyl ether and then dried under vacuum overnight. The final pure white powder was kept in a glovebox for further use.

**X-ray Diffraction (XRD).** XRD analysis was performed on a PANanalytical Empyrean X-ray diffractometer, equipped with a 1.8 kW Cu Kα ceramic X-ray tube (λ = 1.5406 Å) and a PIXcel3D 2 × 2 area detector, operating at 45 kV and 40 mA. NC solutions were concentrated under a flow of nitrogen, then they were drop-cast on a zero-diffraction single crystal substrate. The XRD patterns were then collected under ambient conditions.

**Transmission Electron Microscopy (TEM).** Bright-field TEM (BF-TEM) images with a large field of view were acquired on a JEOL JEM-1400Plus microscope with a thermionic gun ($LaB_6$ crystal), operated at an acceleration voltage of 120 kV. High resolution scanning transmission electron microscopy (HR-STEM) images were acquired on a probe-corrected ThermoFisher Spectra 300 S/TEM operated at 300 kV. Compositional maps were acquired using rapid rastered scanning in Velox, with a probe current of ∼150 pA. Elemental maps were produced after rebinning and local averaging within Velox. NC solutions were diluted ten times in anhydrous toluene and then drop-cast onto copper TEM grids with an ultrathin carbon film.

**Scanning Electron Microscopy-Energy Dispersive X-ray Spectroscopy (SEM-EDS).** SEM analysis was carried out on a JEOL JSM-6490LA scanning electron microscope with thermionic source (W filament) at 30 kV acceleration voltage. Elemental analysis was obtained by means of Energy Dispersive X-ray Spectroscopy (EDS) using a 10 $mm^2$ area Si:Li detector JED-2300 EX-54165JNH from JEOL. The quantification was performed using ZAF method. SEM-EDX analysis was carried out by dropping the sample's dispersion on the silicon substrate, evaporating solvent (toluene) under ambient conditions.

**Optical Measurements.** The UV–visible absorption spectra were recorded using a Varian Cary 300 UV–vis absorption spectrophotometer. The PL spectra of NCs were measured on a Varian Cary Eclipse spectrophotometer ($λ_{ex}$ = 350 nm). The photoluminescence (PL) quantum yield (QY) was measured using a FLS920 Edinburgh Instruments spectrofluorimeter equipped with an integrating sphere. The NC samples were dispersed in anhydrous toluene with an optical density of 0.12 at 400 nm, which was the excitation wavelength employed for PLQY measurements, to minimize self-absorption. The PLQY was measured using a calibrated integrating sphere.

**NMR material and methods.** All the NMR spectra were acquired at 25 °C on a Bruker AvanceIII spectrometer, equipped with a 5 mm QCI cryoprobe. Variable temperature was instead performed at 25, 50 and 80 °C on a Bruker AvanceIII spectrometer, fit with a 5 mm TBO probe. The NMR NC samples were acquired by dispersing the NCs in 600 μL of toluene-D with the use of 5 mm disposable SampleJet (Bruker) tubes. The NMR analyses for ligand quantification were performed using 3 mm disposable SampleJet (Bruker) tubes filled with 200 μL of deuterated dimethyl sulfoxide (DMSO-D), into which the NCs were dissolved. Before the NMR analysis, matching, tuning and spectral resolution were adjusted automatically on each sample-tubes, as well as the 90° optimization.[41]

For $^1H$ NMR spectra 16-128 transients (based on the sample concentration) were gathered without dummy scans, with the receiver fixed at 18, 65536 complex points for FID (Free Induction Decay) and a relaxation delay of 30 sec, on a spectral width of 20.83 ppm, with the offset positioned at 6.18 ppm. A smoothing exponential function equivalent to 0.3 Hz was applied to FIDs prior to the Fourier transform.

The $^1H$-$^1H$ NOESY (Nuclear Overhauser Effect Spectroscopy) experiment (noesygpphpp, Bruker's library) was acquired with 24-64 scans (depending on sample concentration), after 32 steady ones, with 2048 digit points, 256 increments and a mixing time of 300 ms, over a spectral width of 15.15 ppm (offset positioned at 7.49 ppm).

The $^1H$–$^{13}C$ HSQC (Heteronuclear Single-Quantum Coherence Spectroscopy) spectrum (hsqcetgppsi2, Bruker's library) was run by accumulating 16-256 scans (based on sample concentration), after 8 dummy ones, with 2048 complex point, 256 increments and a $^1J_{CH}$ long range of 145 Hz, on a spectral width of 15.15 ppm and 165 ppm, with the offset at 6.69 and 75 ppm for $^1H$ and $^{13}C$ respectively.

$^1H$–$^{13}C$ HMBC (Heteronuclear Multiple Bond Correlation) experiment (hmbcgplpndqf, Bruker's library) was carried out with 32-128 transients (depending on sample concentration), 16 dummy scans, 2048 digit points and 256 increments, by using a $^1J_{CH}$ long range of 10 Hz, over a spectra width of 15.15 and 220 ppm, with the offset at 7 and 100 ppm for $^1H$ and $^{13}C$, respectively.

All the spectra were referred to not deuterated residual solvent peak according to the values for $^1H$ and $^{13}C$ reported in a literature.[42]

To prepare the NC samples for the NMR analyses, we performed a further cleaning cycle, with respect to that described in the synthesis sections above. In details, each NC sample was washed three extra times by precipitation via the addition of 1.5 mL of methyl acetate and redispersion in 0.5 mL of anhydrous toluene. The precipitate was dissolved in 0.6 mL toluene-D for NMR measurement.

**Computational Modeling.** To evaluate the ligand binding affinity of a CsBr ion pair, a tertiary ammonium-Br ion pair and a neutral tertiary amine, we have carried out atomistic simulations at the density functional theory (DFT) level using the PBE exchange–correlation functional[43] and a double-ζ basis set plus polarization functions (DZVP)[44, 45] on all atoms as implemented in the CP2K 6.1[46]. All structures have been optimized in vacuum. Scalar relativistic effects were incorporated as effective core potential functions in the basis set. Spin–orbit coupling effects were not included, but their impact on the relaxed structural properties was demonstrated to be negligible for similar systems. More details on how the models were built can be found in the main text.

To compare the affinity of the tertiary ammonium-Br ion pair on different facet types, we additionally constructed two truncated NC models. In one model, we simply cut the 2.4 nm-sided cubic CsPbBr$_3$ NC model employed previously along the (111) and (-1-1-1) planes, therefore exposing six (001) type facets and two (111) type facets. As cut, this model presented an excess of positive charge that was balanced by randomly removing some Cs$^+$ cations from the edges to obtain a Cs$_{192}$Pb$_{117}$Br$_{426}$ stoichiometry. As for the second model, we truncated the 2.4 nm-sided cubic CsPbBr$_3$ NC model employed previously along the (110), (1-10), (-110) and (-1-10) planes therefore exposing two (001) type facets and four (110) type facets and adjusted the number of Cs$^+$ cations to reach a charge-balanced Cs$_{120}$Pb$_{84}$Br$_{288}$ stoichiometry. Finally, we anchored our simplified version of DDMA$^+$ ligands, i.e. diethyl methyl ammonium, on (*i*) both the (001) facet and the (111) facet of the first model and (*ii*) both the (001) facet and the (110) facet of the second model and compared the total energies of the relaxed structures. All structures have been optimized in vacuum by employing the DFT/PBE/DZVP level of theory. The same approach was followed to compute the affinity of primary and secondary ammonium-Br ion pairs.

The binding energies between the various cations studied in this work and oleate ions were modelled using shorter alkyl chain ligands, i.e. acetate ions. All M(acetate)$_x$ molecular complexes structures have been relaxed in vacuum, again at the DFT/PBE/DZVP level of theory.

## ASSOCIATED CONTENT

**Supporting Information**. Experimental details, stability tests, estimate of the reaction yield, NMR spectra, XRD patterns, SEM- and STEM-EDS, size of M-NCs, sketch of the binding sites and M(acetate)$_x$ complexes.

## AUTHOR INFORMATION

Corresponding Author

* Ivan Infante (ivan.infante@bcmaterials.net)
* Luca de Trizio (luca.detrizio@iit.it)
* Liberato Manna (liberato.manna@iit.it)

Present Addresses

†M.I.: Department of Electrical and Computer Engineering University of Toronto 10 King's College Road, Toronto, Ontario M5S 3G4, Canada.

Author Contributions

The manuscript was written through contributions of all authors.

Funding Sources

J.Z., I.R.P. and L.M. acknowledges funding from the Project IEMAP (Italian Energy Materials Acceleration Platform) within the Italian Research Program ENEA-MASE (Ministero dell'Ambiente e della Sicurezza Energetica) 2021-2024 "Mission Innovation" (agreement 21A033302 GU n. 133/5-6-2021). L.M. acknowledges funding from European Research Council through the ERC Advanced Grant NEHA (grant agreement n. 101095974).

## ACKNOWLEDGMENT

We acknowledge Luca Leoncino for his assistance with the SEM-EDS measurements and Alessandro Migliarino for help with syntheses and discussion. The computing resources and the related technical support used for this work have been provided by CRESCO/ENEAGRID High Performance Computing infra- structure and its staff.[47] CRESCO/ENEAGRID High Performance Computing infrastructure is funded by ENEA, the Italian National Agency for New Technologies, Energy and Sustainable Economic Development and by Italian and European research programmes (http://www.cresco.enea.it/english)

## REFERENCES

(1) Protesescu, L.; Yakunin, S.; Bodnarchuk, M. I.; Krieg, F.; Caputo, R.; Hendon, C. H.; Yang, R. X.; Walsh, A.; Kovalenko, M. V. Nanocrystals of Cesium Lead Halide Perovskites (CsPbX$_3$, X = Cl, Br, and I): Novel Optoelectronic Materials Showing Bright Emission with Wide Color Gamut. *Nano Lett.* **2015**, *15* (6), 3692–3696.
(2) Schmidt, L. C.; Pertegas, A.; Gonzalez-Carrero, S.; Malinkiewicz, O.; Agouram, S.; Minguez Espallargas, G.; Bolink, H. J.; Galian, R. E.; Perez-Prieto, J. Nontemplate Synthesis of CH$_3$NH$_3$PbBr$_3$ Perovskite Nanoparticles. *J. Am. Chem. Soc.* **2014**, *136* (3), 850–853.
(3) Akkerman, Q. A.; D'Innocenzo, V.; Accornero, S.; Scarpellini, A.; Petrozza, A.; Prato, M.; Manna, L. Tuning the Optical Properties of Cesium Lead Halide Perovskite Nanocrystals by Anion Exchange Reactions. *J. Am. Chem. Soc.* **2015**, *137* (32), 10276–10281.
(4) Nedelcu, G.; Protesescu, L.; Yakunin, S.; Bodnarchuk, M. I.; Grotevent, M. J.; Kovalenko, M. V. Fast Anion-Exchange in Highly

Luminescent Nanocrystals of Cesium Lead Halide Perovskites (CsPbX$_3$, X = Cl, Br, I). *Nano Lett.* **2015**, *15* (8), 5635–5640.

(5) Wu, H.; Yang, Y.; Zhou, D.; Li, K.; Yu, J.; Han, J.; Li, Z.; Long, Z.; Ma, J.; Qiu, J. Rb$^+$ Cations Enable the Change of Luminescence Properties in Perovskite (Rb$_x$Cs$_{1-x}$PbBr3) Quantum Dots. *Nanoscale* **2018**, *10* (7), 3429–3437.

(6) Dong, Y.; Qiao, T.; Kim, D.; Parobek, D.; Rossi, D.; Son, D. H. Precise Control of Quantum Confinement in Cesium Lead Halide Perovskite Quantum Dots via Thermodynamic Equilibrium. *Nano Lett.* **2018**, *18* (6), 3716–3722.

(7) Zhang, D.; Eaton, S. W.; Yu, Y.; Dou, L.; Yang, P. Solution-Phase Synthesis of Cesium Lead Halide Perovskite Nanowires. *J. Am. Chem. Soc.* **2015**, *137* (29), 9230–9233.

(8) Ye, S.; Zhao, M.; Song, J.; Qu, J. Controllable Emission Bands and Morphologies of High-Quality CsPbX$_3$ Perovskite Nanocrystals Prepared in Octane. *Nano Res.* **2018**, *11* (9), 4654–4663.

(9) Bera, S.; Behera, R. K.; Pradhan, N. alpha-Halo Ketone for Polyhedral Perovskite Nanocrystals: Evolutions, Shape Conversions, Ligand Chemistry, and Self-Assembly *J. Am. Chem. Soc.* **2020**, *142* (49), 20865–20874.

(10) Almeida, G.; Goldoni, L.; Akkerman, Q.; Dang, Z.; Khan, A. H.; Marras, S.; Moreels, I.; Manna, L. Role of Acid-Base Equilibria in the Size, Shape, and Phase Control of Cesium Lead Bromide Nanocrystals. *ACS Nano* **2018**, *12* (2), 1704–1711.

(11) Shamsi, J.; Raino, G.; Kovalenko, M. V.; Stranks, S. D. To Nano or Not To Nano for Bright Halide Perovskite Emitters. *Nat. Nanotechnol.* **2021**, *16* (11), 1164–1168.

(12) Zhou, Z.-R.; Liao, Z.-H.; Wang, F. Shape-controlled Synthesis of One-Dimensional Cesium Lead Halide Perovskite Nanocrystals: Methods and Advances. *J. Mater. Chem. C* **2023**, *11* (10), 3409–3427.

(13) Ghimire, S.; Chouhan, L.; Takano, Y.; Takahashi, K.; Nakamura, T.; Yuyama, K.-i.; Biju, V. Amplified and Multicolor Emission from Films and Interfacial Layers of Lead Halide Perovskite Nanocrystals. *ACS Energy Lett.* **2018**, *4* (1), 133–141.

(14) Chen, Y.; Smock, S. R.; Flintgruber, A. H.; Perras, F. A.; Brutchey, R. L.; Rossini, A. J. Surface Termination of CsPbBr$_3$ Perovskite Quantum Dots Determined by Solid-State NMR Spectroscopy. *J. Am. Chem. Soc.* **2020**, *142* (13), 6117–6127.

(15) Almeida, G.; Ashton, O. J.; Goldoni, L.; Maggioni, D.; Petralanda, U.; Mishra, N.; Akkerman, Q. A.; Infante, I.; Snaith, H. J.; Manna, L. The Phosphine Oxide Route toward Lead Halide Perovskite Nanocrystals. *J. Am. Chem. Soc.* **2018**, *140* (44), 14878–14886.

(16) Imran, M.; Ijaz, P.; Baranov, D.; Goldoni, L.; Petralanda, U.; Akkerman, Q.; Abdelhady, A. L.; Prato, M.; Bianchini, P.; Infante, I.; Manna, L. Shape-Pure, Nearly Monodispersed CsPbBr$_3$ Nanocubes Prepared Using Secondary Aliphatic Amines. *Nano Lett.* **2018**, *18* (12), 7822–7831.

(17) De Trizio, L.; Infante, I.; Manna, L. Surface Chemistry of Lead Halide Perovskite Colloidal Nanocrystals. *Acc. Chem. Res.* **2023**, *56* (13), 1815–1825.

(18) Creutz, S. E.; Crites, E. N.; De Siena, M. C.; Gamelin, D. R. Anion Exchange in Cesium Lead Halide Perovskite Nanocrystals and Thin Films Using Trimethylsilyl Halide Reagents. *Chem Mater.* **2018**, *30* (15), 4887–4891.

(19) Imran, M.; Caligiuri, V.; Wang, M.; Goldoni, L.; Prato, M.; Krahne, R.; De Trizio, L.; Manna, L. Benzoyl Halides as Alternative Precursors for the Colloidal Synthesis of Lead-Based Halide Perovskite Nanocrystals. *J. Am. Chem. Soc.* **2018**, *140* (7), 2656–2664.

(20) Shamsi, J.; Urban, A. S.; Imran, M.; De Trizio, L.; Manna, L. Metal Halide Perovskite Nanocrystals: Synthesis, Post-Synthesis Modifications, and Their Optical Properties. *Chem. Rev.* **2019**, *119* (5), 3296–3348.

(21) Bunting, J. W.; Thong, K. M. Stability Constants for Some 1 : 1 Metal-Carboxylate Complexes. *Can. J. Chem.* **1970**, *48*, 1654–1656.

(22) Cannan, R. K.; Kibrick, A. Complex Formation between Carboxylic Acids and Divalent Metal Cations. *J. Am. Chem. Soc.* **1938**, *60* (10), 2314–2320.

(23) Robertson, E. J.; Beaman, D. K.; Richmond, G. L. Designated Drivers: The Differing Roles of Divalent Metal Ions in Surfactant Adsorption at the Oil-Water Interface. *Langmuir* **2013**, *29* (50), 15511–15520.

(24) Fromm, K. M. Chemistry of Alkaline Earth Metals: It Is Not All Ionic and Definitely Not Boring! *Coord. Chem. Rev.* **2020**, *408*, 213193.

(25) Stelmakh, A.; Aebli, M.; Baumketner, A.; Kovalenko, M. V. On the Mechanism of Alkylammonium Ligands Binding to the Surface of CsPbBr$_3$ Nanocrystals. *Chem Mater.* **2021**, *33* (15), 5962–5973.

(26) Zaccaria, F.; Zhang, B.; Goldoni, L.; Imran, M.; Zito, J.; van Beek, B.; Lauciello, S.; De Trizio, L.; Manna, L.; Infante, I. The Reactivity of CsPbBr$_3$ Nanocrystals toward Acid/Base Ligands. *ACS Nano* **2022**. 16 (1), 1444–1455.

(27) Maes, J.; Balcaen, L.; Drijvers, E.; Zhao, Q.; De Roo, J.; Vantomme, A.; Vanhaecke, F.; Geiregat, P.; Hens, Z. Light Absorption Coefficient of CsPbBr$_3$ Perovskite Nanocrystals. *J. Phys. Chem. Lett.* **2018**, *9* (11), 3093–3097.

(28) Imran, M.; Ijaz, P.; Goldoni, L.; Maggioni, D.; Petralanda, U.; Prato, M.; Almeida, G.; Infante, I.; Manna, L. Simultaneous Cationic and Anionic Ligand Exchange for Colloidally Stable CsPbBr$_3$ Nanocrystals. *ACS Energy Lett.* **2019**, *4* (4), 819–824.

(29) Shamsi, J.; Abdelhady, A. L.; Accornero, S.; Arciniegas, M.; Goldoni, L.; Kandada, A. R. S.; Petrozza, A.; Manna, L. N-Methylformamide as a Source of Methylammonium Ions in the Synthesis of Lead Halide Perovskite Nanocrystals and Bulk Crystals. *ACS Energy Lett.* **2016**, *1* (5), 1042–1048.

(30) Montgomery, C. D. Factors Affecting Energy Barriers for Pyramidal Inversion in Amines and Phosphines: A Computational Chemistry Lab Exercise. *J. Chem. Educ.* **2013**, *90* (5), 661–664.

(31) Ravi, V. K.; Santra, P. K.; Joshi, N.; Chugh, J.; Singh, S. K.; Rensmo, H.; Ghosh, P.; Nag, A. Origin of the Substitution Mechanism for the Binding of Organic Ligands on the Surface of CsPbBr$_3$ Perovskite Nanocubes. *J. Phys. Chem. Lett.* **2017**, *8* (20), 4988–4994.

(32) Quarta, D.; Imran, M.; Capodilupo, A. L.; Petralanda, U.; van Beek, B.; De Angelis, F.; Manna, L.; Infante, I.; De Trizio, L.; Giansante, C. Stable Ligand Coordination at the Surface of Colloidal

CsPbBr$_3$ Nanocrystals. *J. Phys. Chem. Lett.* **2019**, *10* (13), 3715−3726.
(33) Ding, Y.; Zhang, Z.; Toso, S.; Gushchina, I.; Trepalin, V.; Shi, K.; Peng, J. W.; Kuno, M. Mixed Ligand Passivation as the Origin of Near-Unity Emission Quantum Yields in CsPbBr$_3$ Nanocrystals. *J. Am. Chem. Soc.* **2023**, *145* (11), 6362−6370.
(34) Almeida, G.; Ashton, O. J.; Goldoni, L.; Maggioni, D.; Petralanda, U.; Mishra, N.; Akkerman, Q. A.; Infante, I.; Snaith, H. J.; Manna, L. The Phosphine Oxide Route toward Lead Halide Perovskite Nanocrystals. *J. Am. Chem. Soc.* **2018**, *140* (44), 14878−14886.
(35) ten Brinck, S.; Infante, I. Surface Termination, Morphology, and Bright Photoluminescence of Cesium Lead Halide Perovskite Nanocrystals. *ACS Energy Lett.* **2016**, *1* (6), 1266−1272.
(36) Bunting, J. W.; Thong, K. M. Stability Constants for Some 1:1 Metal−carboxylate Complexes. *Can. J. Chem.* **1970**, *48* (11), 1654−1656.
(37) Robertson, E. J.; Beaman, D. K.; Richmond, G. L. Designated Drivers: The Differing Roles of Divalent Metal Ions in Surfactant Adsorption at the Oil−Water Interface. *Langmuir* **2013**, *29* (50), 15511−15520.
(38) Mendes de Oliveira, D.; Zukowski, S. R.; Palivec, V.; Hénin, J.; Martinez-Seara, H.; Ben-Amotz, D.; Jungwirth, P.; Duboué-Dijon, E. Binding of Divalent Cations to Acetate: Molecular Simulations Guided by Raman Spectroscopy. *Phys. Chem. Chem. Phys.* **2020**, *22* (41), 24014−24027.
(39) Weidman, M. C.; Seitz, M.; Stranks, S. D.; Tisdale, W. A. Highly Tunable Colloidal Perovskite Nanoplatelets through Variable Cation, Metal, and Halide Composition. *ACS Nano* **2016**, *10* (8), 7830−7839.
(40) Mucci, A.; Domain, R.; Benoit, R. L. Solvent Effect on the Protonation of Some Alkylamines. *Can. J. Chem.* **1980**, *58* (9), 953−958.
(41) Wu, P. S.; Otting, G. Rapid Pulse Length Determination in High-Resolution NMR. *J. Magn. Reson.* **2005**, *176* (1), 115−119.
(42) Fulmer, G. R.; Miller, A. J.; Sherden, N. H.; Gottlieb, H. E.; Nudelman, A.; Stoltz, B. M.; Bercaw, J. E.; Goldberg, K. I. NMR Chemical Shifts of Trace Impurities: Common Laboratory Solvents, Organics, and Gases in Deuterated Solvents Relevant to the Organometallic Chemist. *Organometallics* **2010**, *29* (9), 2176−2179.
(43) Perdew, J. P.; Burke, K.; Ernzerhof, M. Generalized Gradient Approximation Made Simple. *Phys. Rev. Lett.* **1996**, *77* (18), 3865−3868.
(44) Hutter, J.; Iannuzzi, M.; Schiffmann, F.; VandeVondele, J. cp2k: Atomistic Simulations of Condensed Matter Systems. *WIREs Computational Molecular Science* **2014**, *4* (1), 15−25.
(45) VandeVondele, J.; Hutter, J. Gaussian Basis Sets for Accurate Calculations on Molecular Systems in Gas and Condensed Phases. *J. Chem. Phys.* **2007**, *127* (11), 114105.
(46) Kühne, T. D.; Iannuzzi, M.; Del Ben, M.; Rybkin, V. V.; Seewald, P.; Stein, F.; Laino, T.; Khaliullin, R. Z.; Schütt, O.; Schiffmann, F.; et al. CP2K: an Electronic Structure and Molecular Dynamics Software Package - Quickstep: Efficient and Accurate Electronic Structure Calculations. *J. Chem. Phys.* **2020**, *152* (19), 194103.
(47) Iannone, F.; Ambrosino, F.; Bracco, G.; De Rosa, M.; Funel, A.; Guarnieri, G.; Migliori S.; Palombi F.; Ponti, G.; Santom, G.; Procacci, P. CRESCO ENEA HPC Clusters: A Working Example of A Multifabric GPFS Spectrum Scale Layout. *2019 International Conference on High Performance Computing & Simulation (HPCS)* **2019**, Dublin, Ireland 1051−1052.

Supporting Information for:

# Exogenous Metal Cations in the Synthesis of CsPbBr$_3$ Nanocrystals and their Interplay with Tertiary Amines

Zhanzhao Li[1], Luca Goldoni[2], Ye Wu[1], Muhammad Imran[1,†], Yurii P. Ivanov[3], Giorgio Divitini[3], Juliette Zito[1], Iyyappa Rajan Panneerselvam[1], Dmitry Baranov[1,4], Ivan Infante[5,6]*, Luca De Trizio[2]*, Liberato Manna[1]*

[1] Nanochemistry, Istituto Italiano di Tecnologia, Via Morego 30, Genova, Italy

[2] Chemistry Facility, Istituto Italiano di Tecnologia, Via Morego 30, Genova, Italy

[3] Electron Spectroscopy and Nanoscopy, Istituto Italiano di Tecnologia, Via Morego 30, Genova, Italy

[4] Division of Chemical Physics, Department of Chemistry, Lund University, P.O. Box, 124, SE-221 00 Lund, Sweden

[5] BCMaterials, Basque Center for Materials, Applications, and Nanostructures, UPV/EHU Science Park, Leioa 48940, Spain

[6] Ikerbasque Basque Foundation for Science Bilbao 48009, Spain

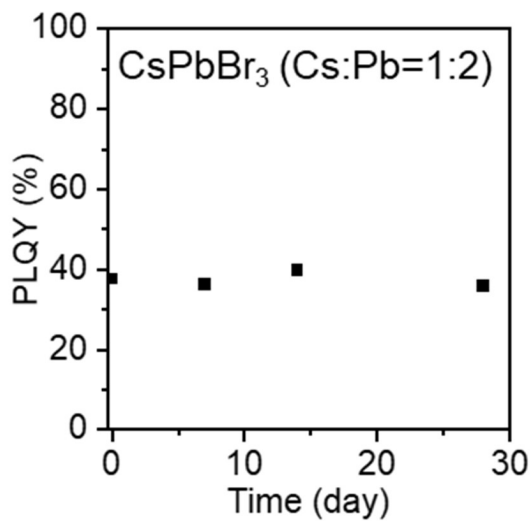

**Figure S1.** PLQY values measured for the CsPbBr$_3$ NC sample of Figure 1a, prepared with the "minimal synthesis" approach (with a Cs:Pb = 1:2 ratio), over a time span of one month.

**Method for estimating the reaction yield** of colloidal CsPbBr$_3$ by optical absorption spectroscopy.[1] Three 20 µL aliquots of a crude reaction mixture containing CsPbBr$_3$ NCs were diluted in 10 ml anhydrous toluene. We then measured the optical density of these three equivalent dispersions. The absorption intensity (A) at 400 nm was considered for the subsequent calculations. We employed the Beer–Lambert law:

$$c = A/\varepsilon \times l$$

where ε is the molar absorptivity of the absorbing species (that is CsPbBr$_3$ NCs), $\varepsilon = (2.42 \pm 0.04) \times 10^{-2} \times d^3$ at 400 nm, d is the size of the NCs and l is the path length, we could calculate the the concentration of the absorbing species (c), and, consequently, the moles (n) of NCs in the crude reaction solution from the volume of the solution. The Pb yield was eventually calculated as Yield$_{Pb}$= n/n$_0$ × 100%, where n$_0$ are the moles of Pb precursor employed in the synthesis.

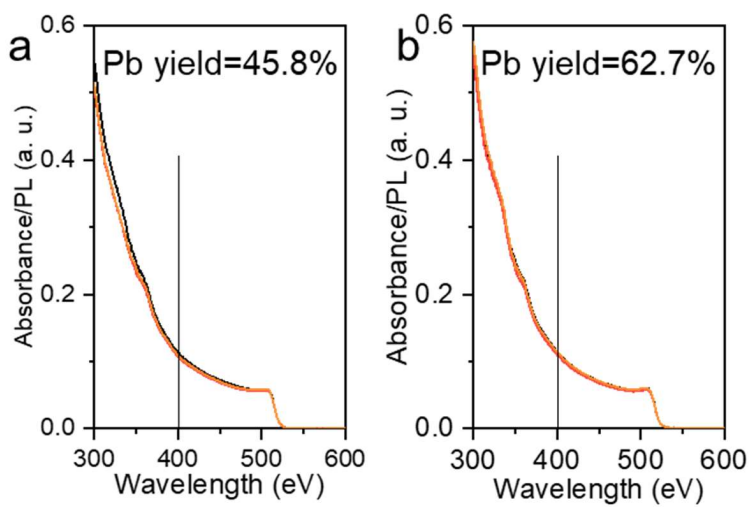

**Figure S2.** Estimate of the reaction yield (in terms of Pb incorporated in the CsPbBr$_3$ NCs) by optical absorption spectroscopy of the crude solution of (a) CsPbBr$_3$ (Cs:Pb=1:2) NCs and (b) CsPbBr$_3$ (Cs:Pb=1:1.5) NCs made with the "minimal synthesis" approach.

S16

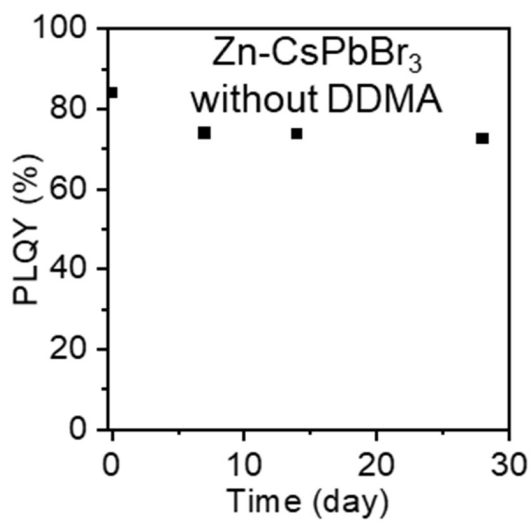

**Figure S3.** PLQY values measured for CsPbBr$_3$ NCs (Figure 2a), made with Cs:Pb:Zn = 1:1:1 in the "minimal synthesis" approach with no addition of DDMA nor DDMA-Br, over a time span of one month.

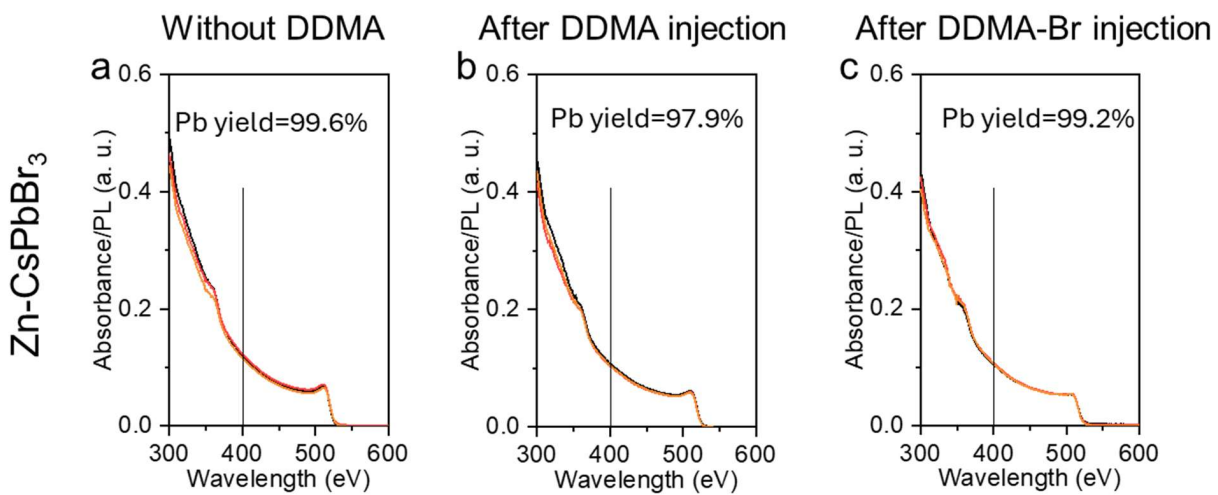

**Figure S4.** Estimation of the Pb yield by optical absorption spectroscopy of the crude solution of CsPbBr$_3$ NCs made employing Zn$^{2+}$ cations in the "minimal synthesis" approach with either (a) no DDMA, (b) DDMA and (c) DDMA-Br. In the last two cases, namely (b) and (c), the reaction was quenched 10 sec after the addition of DDMA or DDMA-Br.

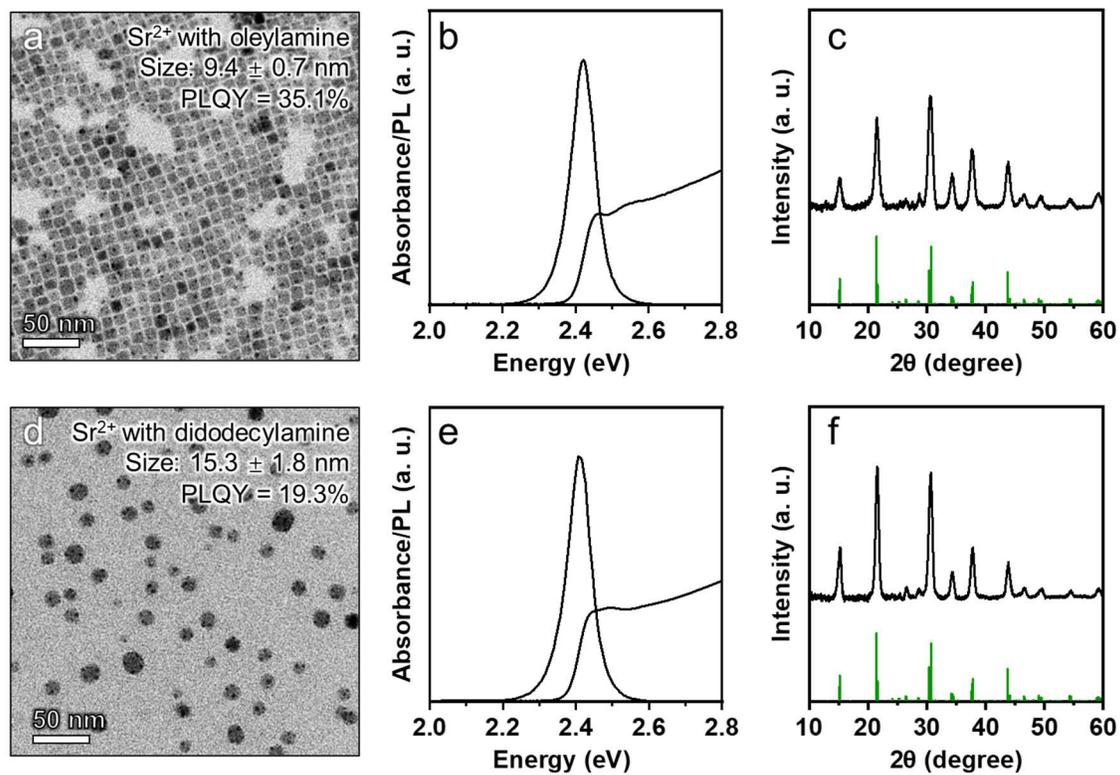

**Figure S5.** (a, d) TEM images, (b, e) optical absorption and photoluminescence spectra and (c, f) XRD patterns of NCs synthesized employing $Sr^{2+}$ as exogenous cation and a primary amine (oleylamine, panels a-c) or a secondary amine (didodecylamine, panels d-f).

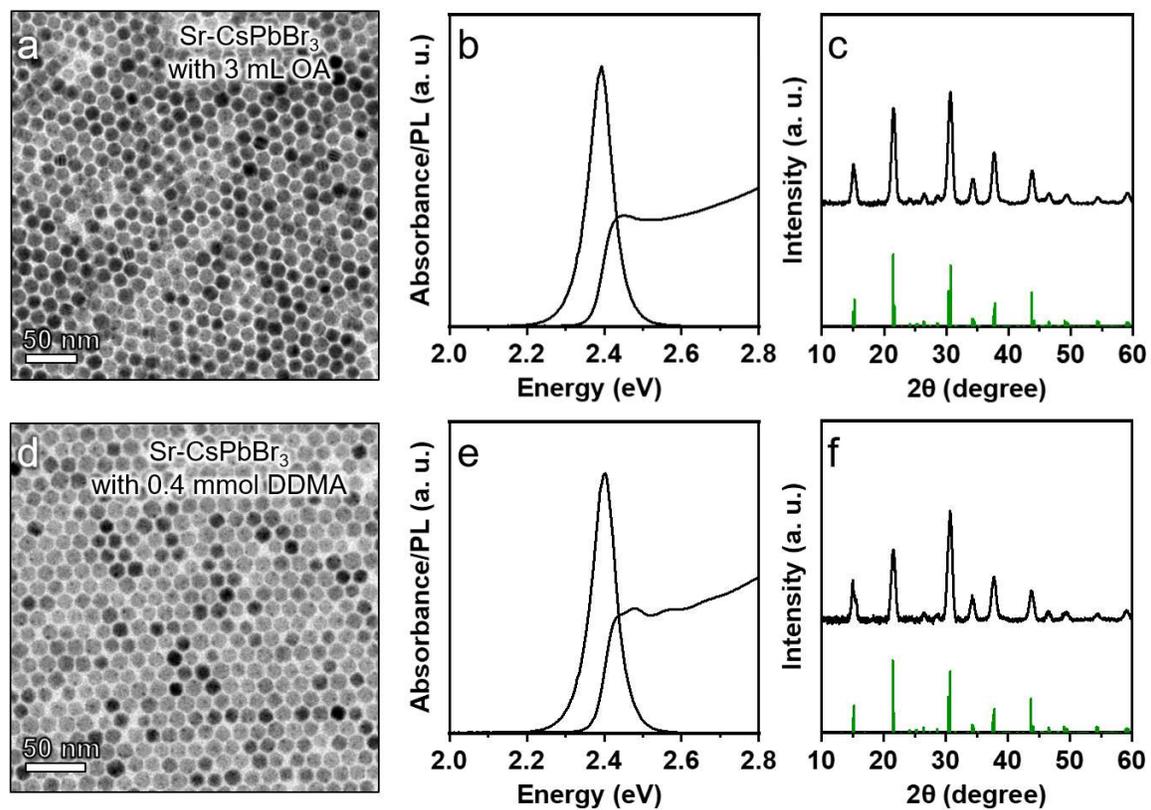

**Figure S6.** (a, d) TEM images, (b, e) optical absorption and photoluminescence spectra and (c, f) XRD patterns of NCs synthesized employing a double amount of either oleic acid (3 mL, panels a-c) or DDMA (0.4mmol, panels d-f).

**Nuclear Magnetic Resonance (NMR)**

*Comparison between the $^1$H NMR spectra of Sr-based CsPbBr$_3$ NCs (made with DDMA) and that of free DDMA, in toluene-D*

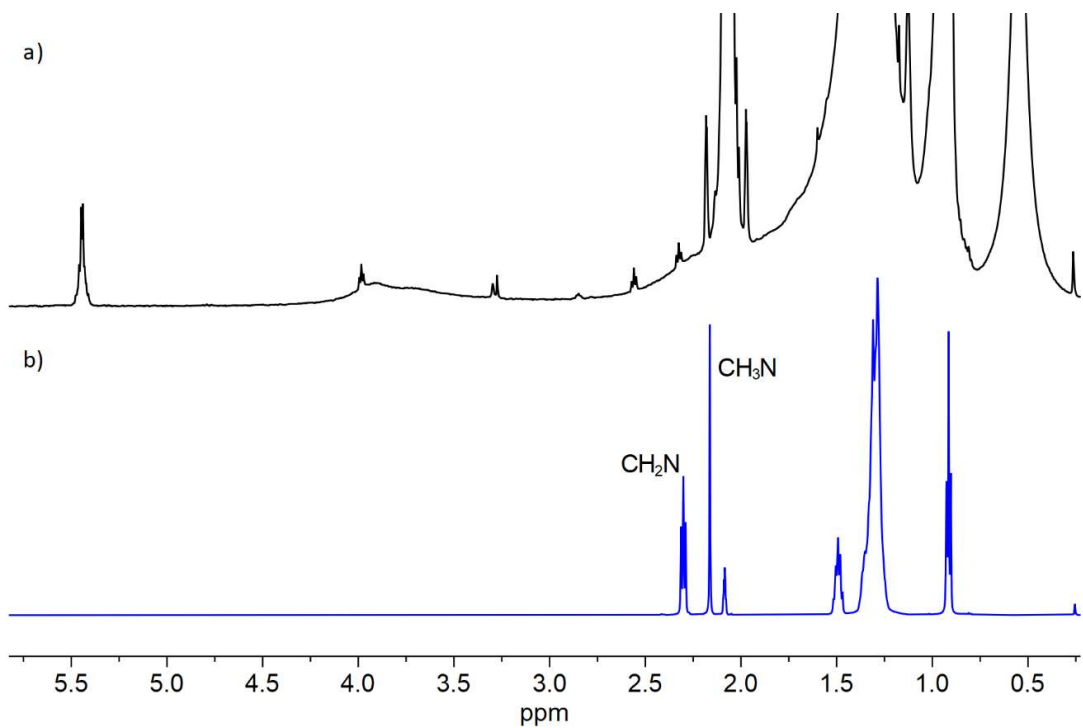

**Figure S7.** $^1$H NMR spectra in toluene-D of a) CsPbBr$_3$ NCs (made with Sr$^{2+}$ and DDMA, reaction quenched after 10 sec) and b) DDMA.

In the $^1$H NMR spectrum of Sr-based NCs (Figure S7a) the diagnostic peaks of DDMA, i.e. the CH$_2$ and CH$_3$ on nitrogen atom, are shifted to lower ppm and broadened with respect to those of the free amine (Figure S7b). This indicates that the amine, in its protonated form (Figure 3), is interacting with the NC's surface, assuming the slowed down tumbling regime, i.e. the correlation time ($\tau_c$) of the NCs in solution.

*2D NMR experiments of the sole DDMA in the same solvent (toluene-D) for comparison*

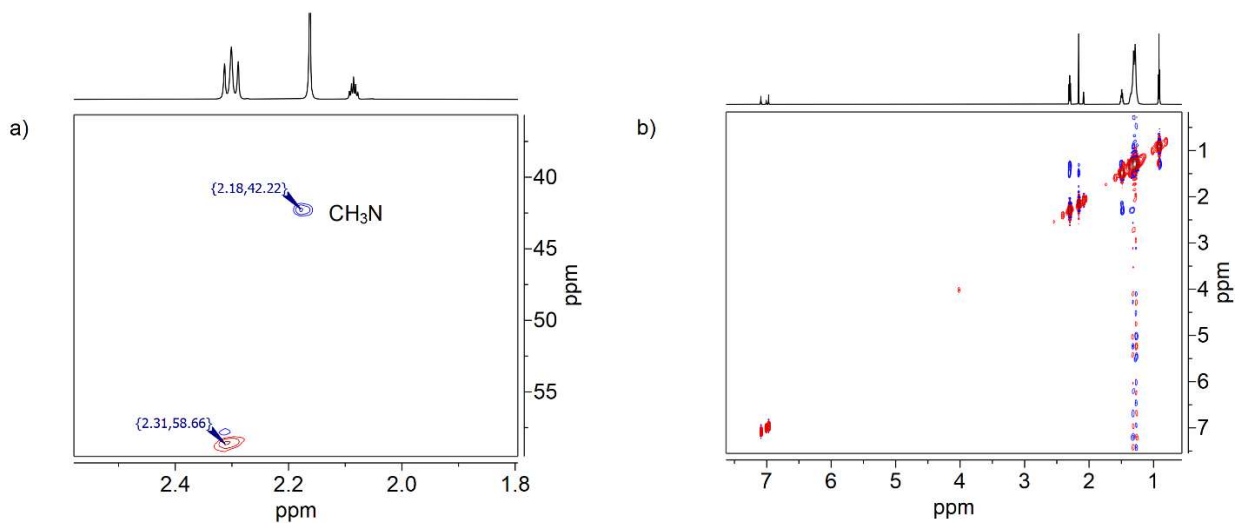

**Figure S8.** a) $^1H–^{13}C$ HSQC and b) $^1H–^1H$ NOESY spectra of pristine DDMA, in toluene-D. The NOE cross peaks are positive (blue) with a correlation time $\tau_c$, typical of small molecule.

*2D NMR spectra of Zn-based NCs in toluene-D*

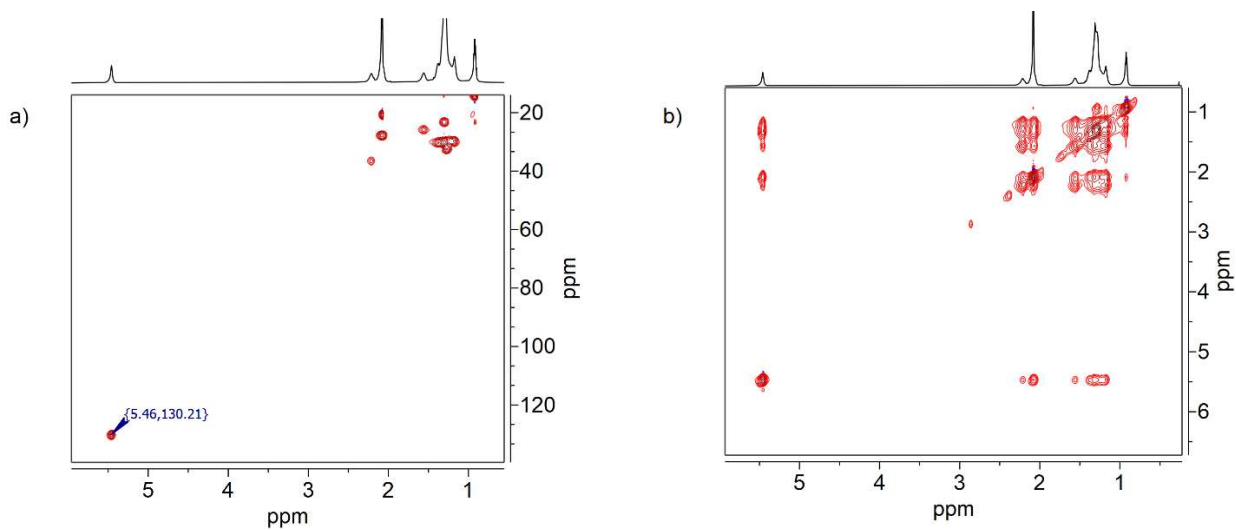

**Figure S9.** a) $^1H–^{13}C$ HSQC spectrum CsPbBr$_3$ NCs (made with Zn$^{2+}$ and DDMA, reaction quenched after 10 sec) in toluene-D. b) The $^1H$-$^1H$ NOESY spectrum in toluene-D returns for oleic acid, negative (red) NOE cross peaks, characteristic of species with a slowed down tumbling regime in solution due to the binding with the NC's surface.

This analysis indicates that CsPbBr$_3$ NCs made with Zn$^{2+}$ and DDMA are passivated only by oleate species and no DDMA$^+$ cations.

*¹H NMR of Sr-based NCs in DMSO-D*

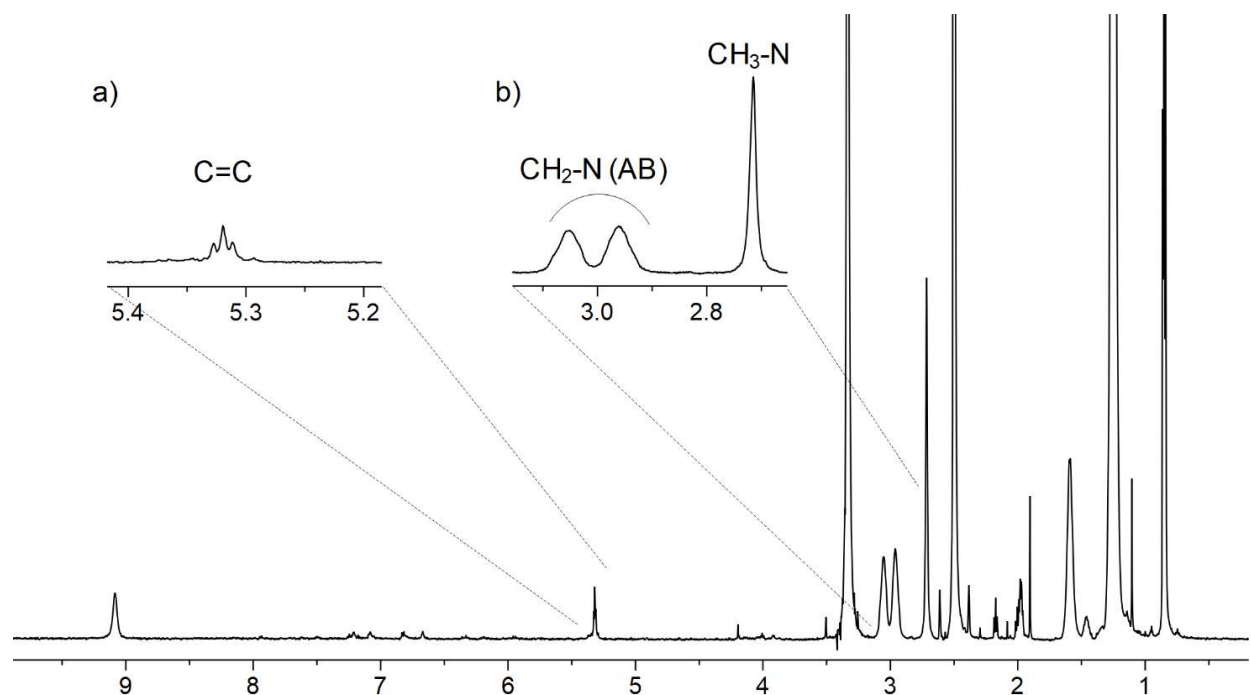

**Figure S10.** ¹H NMR quantitative spectrum of CsPbBr$_3$ NCs (made with Sr$^{2+}$ and DDMA, reaction quenched after 10 sec) dissolved in DMSO-D. The insets are the expanded regions of: a) the diagnostic double bound signal of Oleic Acid (*m*, 5.32 ppm, 2H); b) the AB system (*br m*, 3.05 and 2.96 ppm, 2H) attributed to the nonequivalent protons of CH$_2$N and the signal (*br s*, 2.72 ppm, 3H) attributed to the CH$_3$-N of protonated DDMA.

Both the shift to low field (from 2.09 ppm of the CH$_2$N of free DDMA, see Figure S13, to 2.96 and 3.05 ppm) and the splitting into AB system (see also Figure S11) confirm the presence of protonated DDMA (see also the discussion below related to Figures S12-15).

*2D NMR spectra of Sr-based NCs*

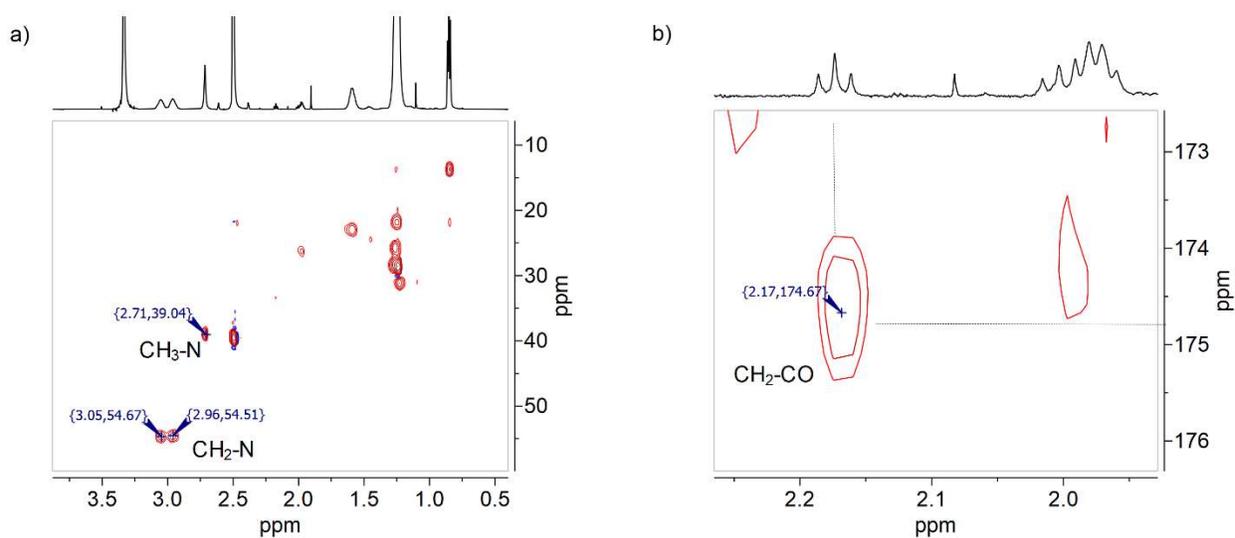

**Figure S11.** a) $^1$H–$^{13}$C HSQC spectrum of CsPbBr$_3$ NCs (made with Sr$^{2+}$ and DDMA, reaction quenched after 10 sec) dissolved in DMSO-D. Notably CH$_2$N resonances are magnetically not equivalent (AB system). Both CH$_2$-N and CH$_3$-N resonances appear downfield shifted both at $^1$H and at $^{13}$C compared to the corresponding values of the sole DDMA (Figure S7A). b) $^1$H–$^{13}$C HMBC of the same NCs dissolved in DMSO-D. The HMBC cross correlation between the CH$_2$ the CO verifies that such signal belongs to Oleic acid.

These analyses indicated that Sr-based CsPbBr$_3$ NCs are passivated by both DDMA$^+$ and oleate species.

*$^1$H NMR and HSQC of Sr-based NCs in DMSO-D in which we also added trifluoro acetic acid to force the DDMA protonation*

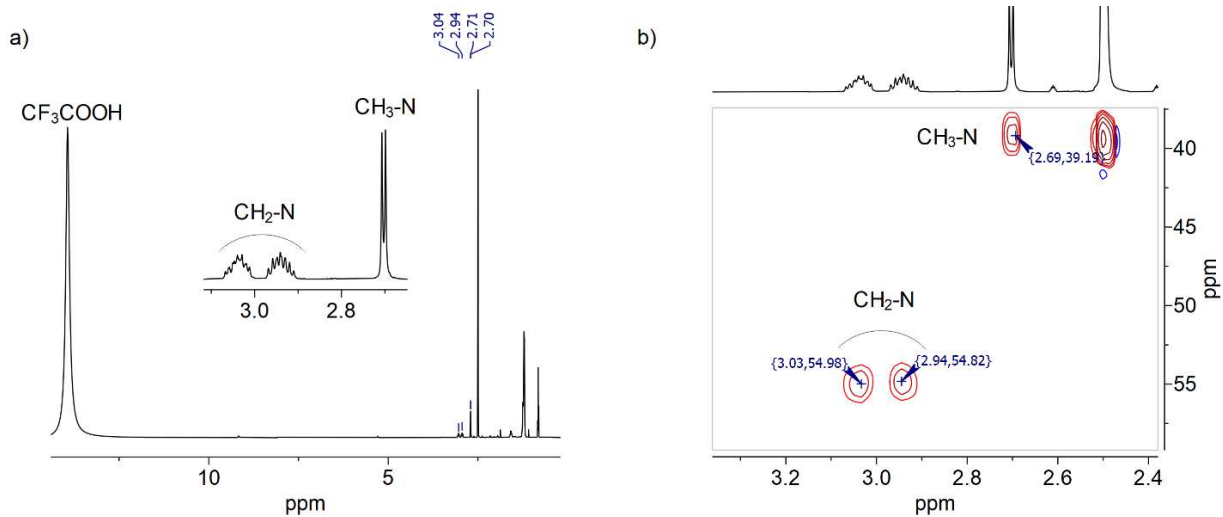

**Figure S12.** a) $^1$H and b) $^1$H–$^{13}$C HSQC spectra of CsPbBr$_3$ NCs (made with Sr$^{2+}$ and DDMA, reaction quenched after 10 sec) dissolved in DMSO-D with a large excess of trifluoro acetic acid (TFA, CF$_3$COOH) aimed at pushing further the protonation of DDMA. Notably CH$_2$N resonances are magnetically not equivalent (AB system). The chemical shift both at $^1$H and $^{13}$C are identical to those reported in Figure S11, further confirming the protonation of DDMA molecules passivating Sr-based CsPbBr$_3$ NCs.

*DDMA characterization in DMSO-D*

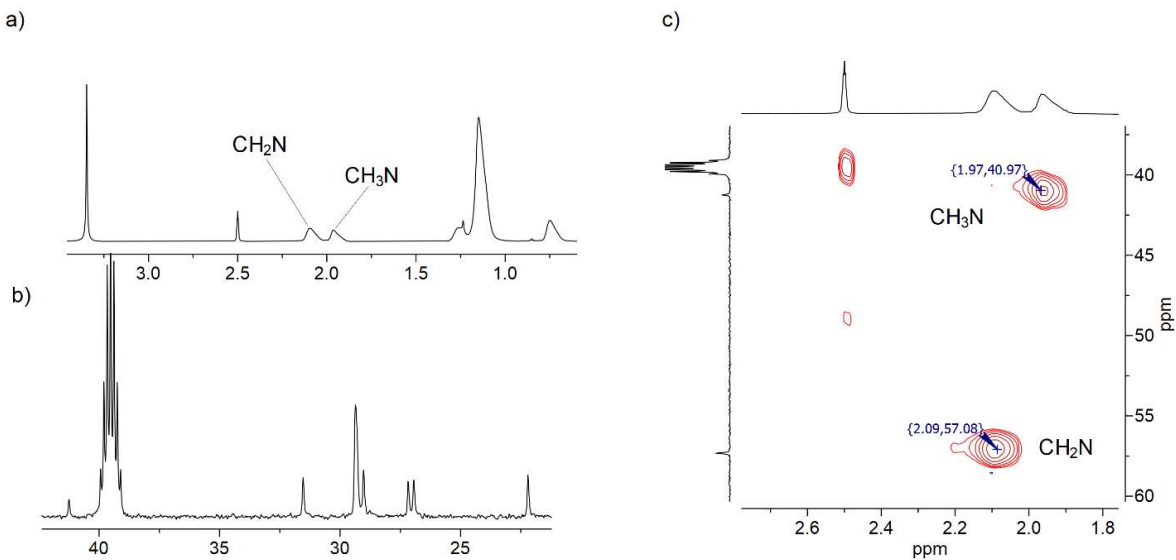

**Figure S13.** a) $^1$H, b) $^{13}$C, and c) $^1$H–$^{13}$C HSQC spectra and diagnostic peaks assignment of DDMA in DMSO-D.

The CH$_2$N protons of neutral DDMA are characterized by a broad peak at 2.09 ppm with the $^{13}$C at 57.08 ppm. The CH$_2$N protons of DDMA are magnetically equivalent by definition, as they exhibit the same chemical shift and coupling constants.[2]

*DDMA characterization in DMSO-D with large excess of TFA*

As a control experiment, we characterized DDMA in the presence of a large excess of trifluoroacetic acid (TFA). The CH$_2$-N signals of the didodecyl groups were low-field shifted, with the $^1$H resonances split into two sets of multiplets at 3.04 and 2.95 ppm (see Figure S14). Both signals correlated with the same carbon at 54.84 ppm in the HSCQ spectrum (Figure S15 a), further confirmed by the direct $^{13}$C spectrum (Figure S15 b). Moreover, both CH$_2$N resonances showed the same connectivity, as indicated by the COSY experiment (Figure S15 c) and the long-distance $^1$H–$^{13}$C HMBC experiment (Figure S15 d). These results indicate that both multiplets correspond to CH$_2$N protons of protonated DDMA. Since they differ in chemical shift in the $^1$H NMR spectrum, these CH$_2$-N protons are not magnetically equivalent by definition.

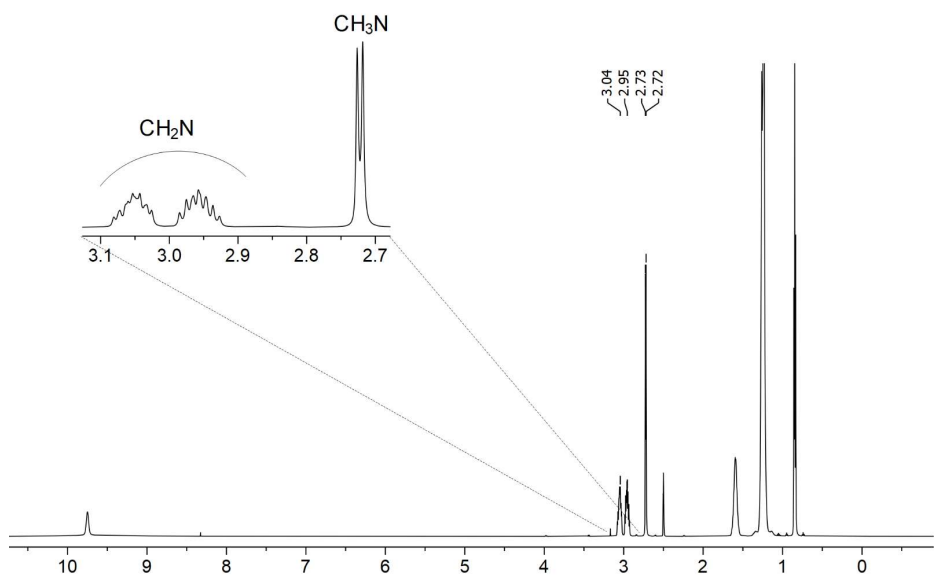

**Figure S14.** $^1$H NMR spectrum and diagnostic peaks assignment of sole DDMA in DMSO-D in presence of a large excess of TFA.

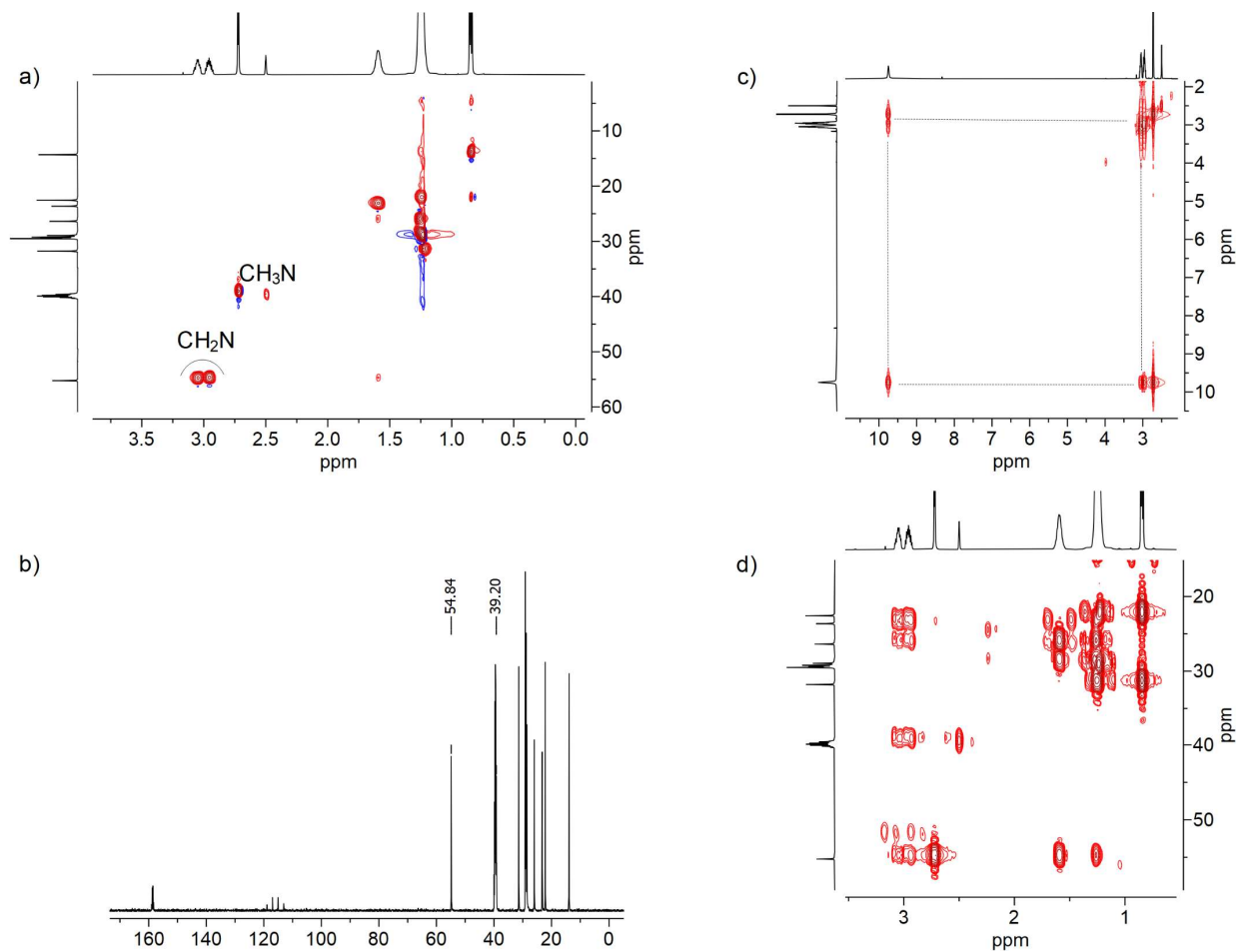

**Figure S15**. a) $^1$H-$^{13}$C HSQC, b) $^{13}$C, c) $^1$H COSY and d) $^1$H–$^{13}$C HMBC NMR spectra and diagnostic peaks assignment of sole DDMA in DMSO-D in presence of a large excess of TFA.

Both the chemical shift and the splitting pattern of the CH$_2$N resonances of protonated DDMA are identical to those observed when CsPbBr$_3$ NCs (prepared with Sr$^{2+}$ and DDMA) are dissolved in DMSO-D (Figure S10 and S11). This indicates that such NCs are passivated by protonated DDMA.

We speculate that the non-equivalence of the CH$_2$N protons of protonated DDMA is due to both the steric hindrance of the substituents, i.e., the didodecyl and methyl groups, and to the fact that the inversion process on nitrogen is possible only for non-protonated amines (as they feature a lone pair in a sp3 orbital). Such inversion at room temperature is favored and rapid on the NMR time scale in the presence of a lone pair because it goes through a trigonal planar intermediate,[3] resulting in a mediated broad signal for the CH$_2$N resonances of DDMA. However, with an excess of TFA (i.e., protonated DDMA), the lone pair is used for the N-H bond, and therefore the inversion process is hindered.

*¹H NMR of Zn-based NCs dissolved in DMSO-D with and without the addition of TFA*

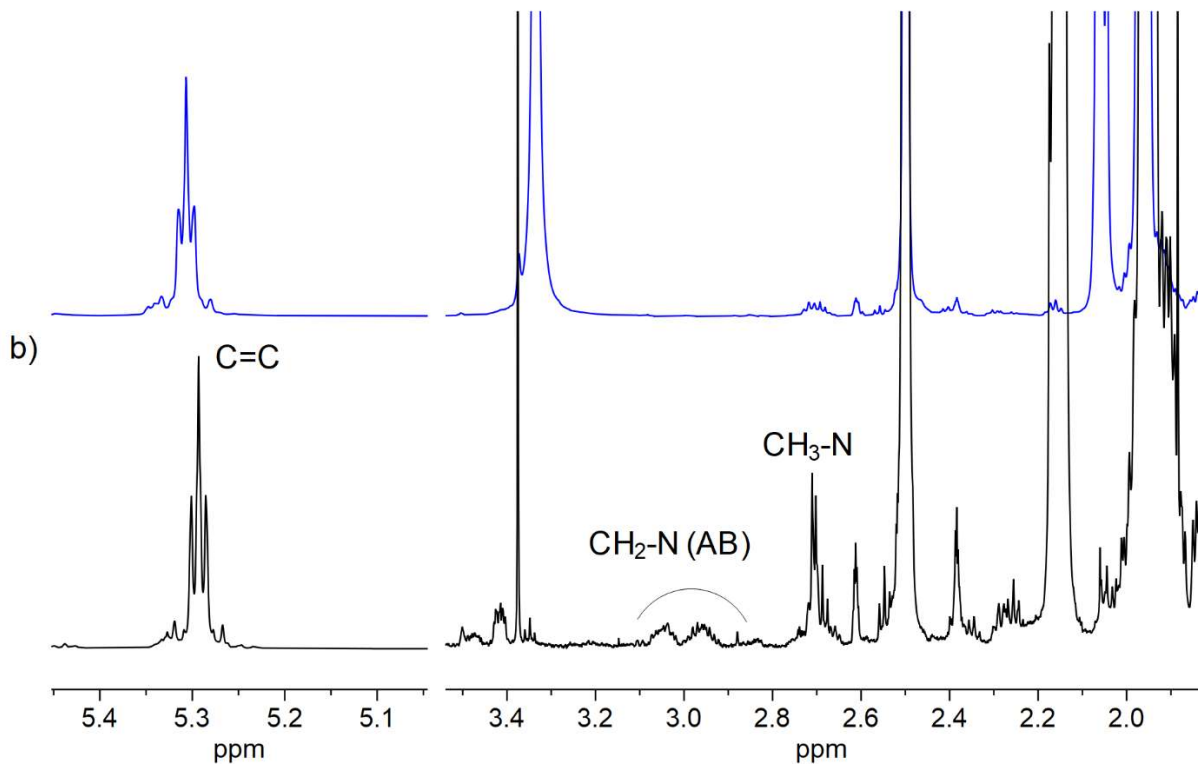

**Figure S16.** ¹H NMR spectra of CsPbBr$_3$ NCs (made with Zn$^{2+}$ and DDMA and quenched after 10 sec) dissolved in a) DMSO-D and b) in DMSO-D with the addition of a large excess of TFA.

In a) the region of protonated amine from 2.7 to 3.1 ppm do not feature any peak (indicating the absence of DDMA$^+$). In b) the addition of TFA results in the appearance of the characteristic peaks of protonated amine (AB system) in the ¹H spectrum. This indicated that DDMA$^+$ ions were not present on the surface of CsPbBr$_3$ NC made with Zn$^{2+}$ and DDMA, but they formed only when forcing the protonation of residual DDMA molecules via TFA addition.

*NMR experiments aimed at understanding the protonation of DDMA in toluene in either the presence or absence of $Zn^{2+}$ and $Sr^{2+}$ cations at different temperatures*

To investigate the protonation of DDMA in toluene we recorded $^1H$ NMR spectra of a mixture of oleic acid and DDMA (see the experimental conditions below) at different temperatures, namely 25, 50 and 80°C (Figure S17, light blue triangles). Our analyses revealed that the protonation of the amine (shift to lower ppm) was favored at higher temperatures. The same experiments performed in the presence of either $Sr^{2+}$ or $Zn^{2+}$ indicated that $Sr^{2+}$ cations systematically enhanced the protonation of the amine (Figure S17, black squares), while $Zn^{2+}$ cations inhibited it (Figure S17, red dots). To confirm that the protonation of DDMA results in a shift to lower ppm we also performed a series of $^1H$ NMR analyses in which DDMA was mixed with increasing amounts of TFA (which, being a strong acid, is known to efficiently protonate amines). These experiments clearly indicate that the protonation of the amine is indeed followed by a shift of $^1H$ signals to lower ppm.

The experimental conditions:
  -*Oleic acid and DDMA (no $Zn^{2+}$ or $Sr^{2+}$ cations):* 2 mmol oleic acid and 0.2 mmol DDMA were mixed inside a nitrogen filled glovebox. 10 μL of this mixture was diluted in 1 ml of anhydrous toluene-D.
  -*$Sr^{2+}$ case:* 2 mmol oleic acid and 0.1 mmol of $Sr(OAc)_2$ were mixed in a vial inside a nitrogen filled glovebox. Then the resulting mixture was heated to 180 °C for 10 min to form the metal oleate complexes. The resulting solution was cooled down to room temperature and mixed with 0.2 mmol of DDMA. 10 μL of this mixture was taken to dilute it in 1 ml of anhydrous toluene-D.
  -*$Zn^{2+}$ case:* 2 mmol oleic acid and 0.1 mmol of $Zn(OAc)_2$ were mixed in a vial inside a nitrogen filled glovebox. Then the resulting mixture was heated to 180 °C for 10 min to form the metal oleate complexes. The resulting solution was cooled down to room temperature and mixed with 0.2 mmol of DDMA. 10 μL of this mixture was taken to dilute it in 1 ml of anhydrous toluene-D.
  -*TFA cases:* 2 mmol oleic acid and 0.2 mmol DDMA were mixed inside a nitrogen filled glovebox. Subsequently, 5, 10 or 15 ul of TFA were added to the mixture. 10 μL of each of these mixtures were diluted in 1 ml of anhydrous toluene-D.

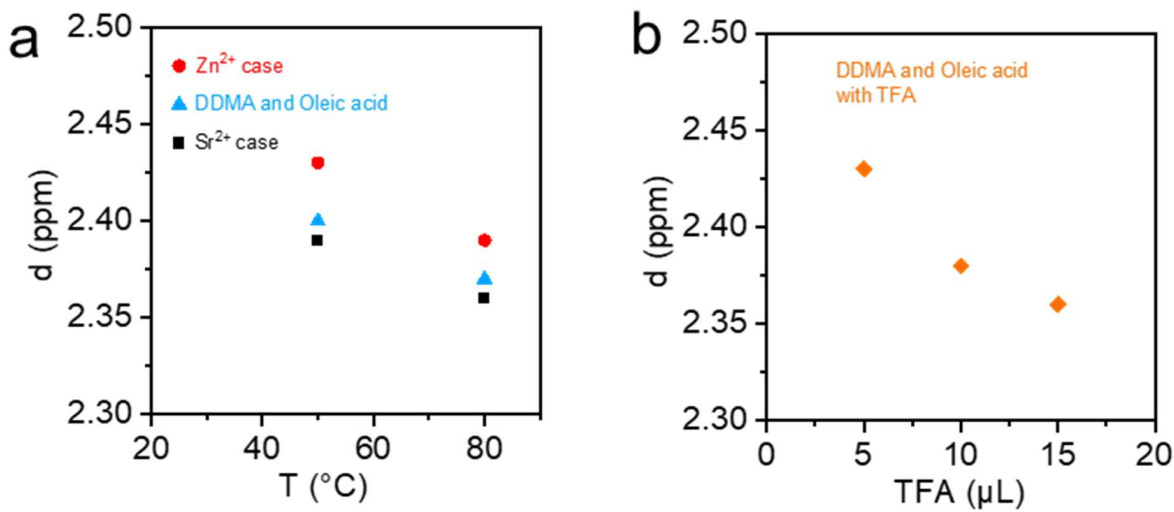

**Figure S17.** a) $^1H$ NMR chemical shift δ (ppm) of $CH_2$-N protons in a mixture of oleic acid and DDMA in toluene-D in absence of exogenous metal cations (light blue triangles) or in the presence of either $Sr^{2+}$ cations

(black squares) or $Zn^{2+}$ cations (red dots) at different temperatures (25, 50 and 80 °C). b) $^1$H NMR chemical shift δ (ppm) of $CH_2$-N protons in a mixture of oleic acid, DDMA and different amounts of TFA (5, 10 or 15 uL) at room temperature.

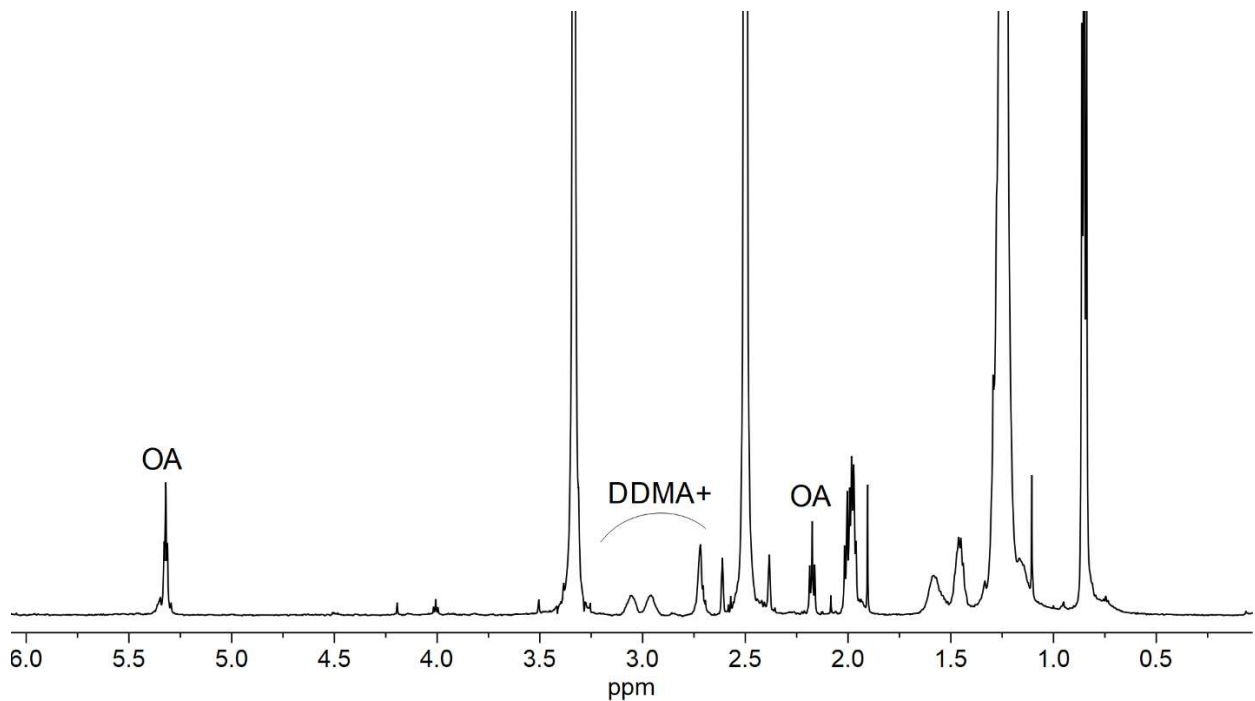

**Figure S18.** $^1$H NMR quantitative spectrum of $CsPbBr_3$ NCs (made with $Zn^{2+}$ and DDMA-Br, reaction quenched after 10 sec) dissolved in DMSO-D.

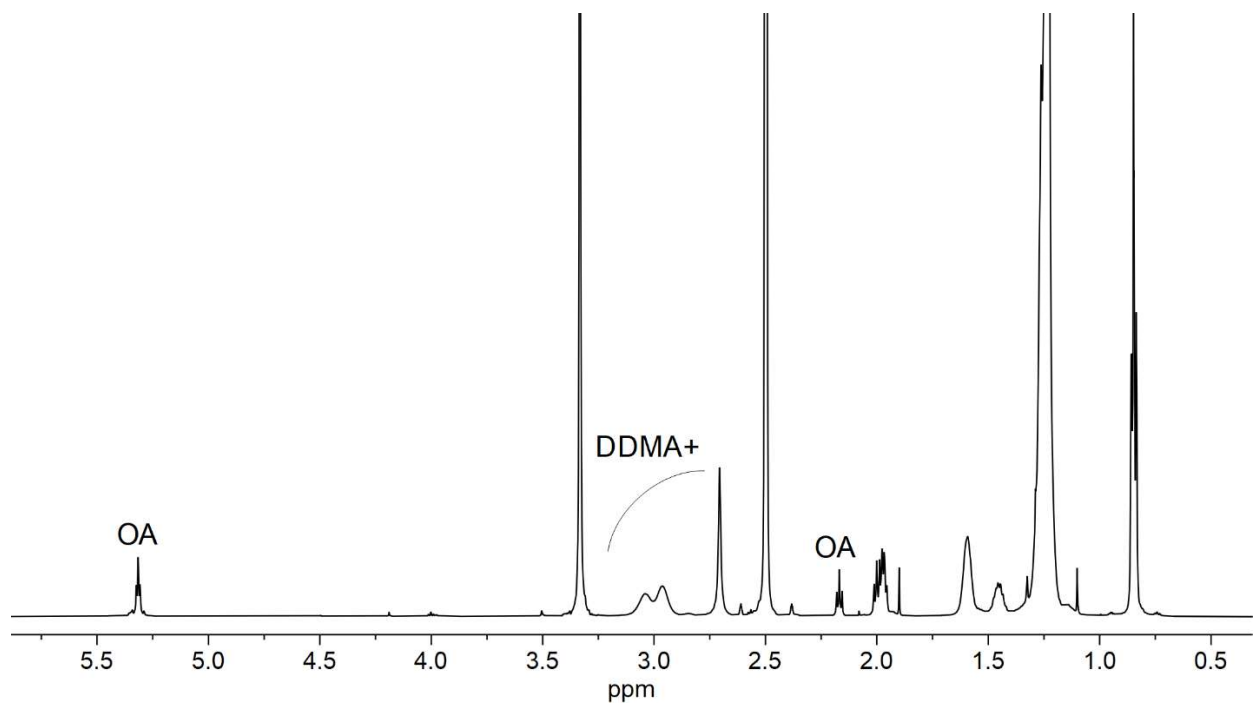

**Figure S19.** $^1$H NMR quantitative spectrum of CsPbBr$_3$ NCs (made with Sr$^{2+}$ and DDMA-Br, reaction quenched after 10 sec) dissolved in DMSO-D.

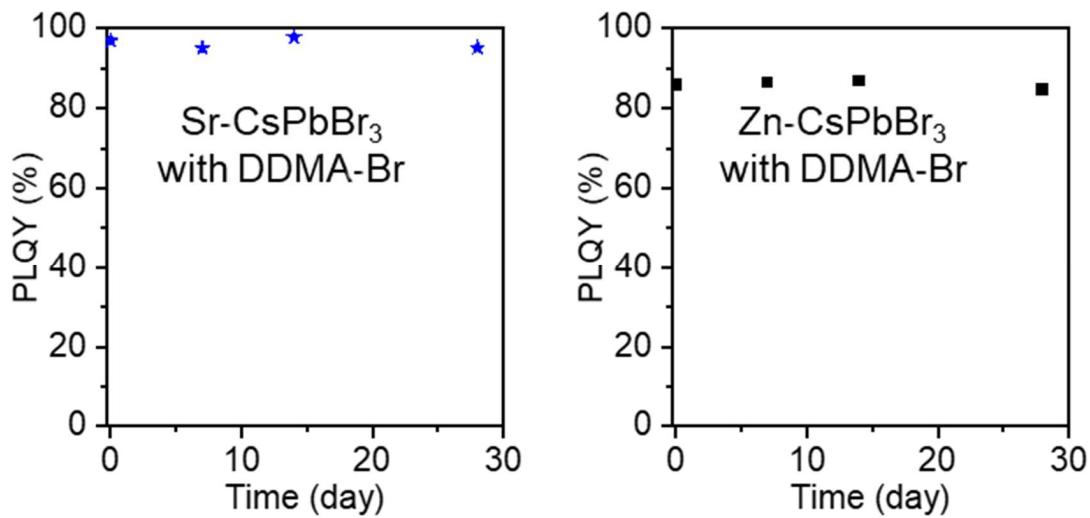

**Figure S20.** PLQY values measured for CsPbBr$_3$ NCs made with (a) Sr$^{2+}$ and DDMA-Br or (b) Zn$^{2+}$ and DDMA-Br (quenched 10 sec after the injection of DDMA-Br), over a time span of one month.

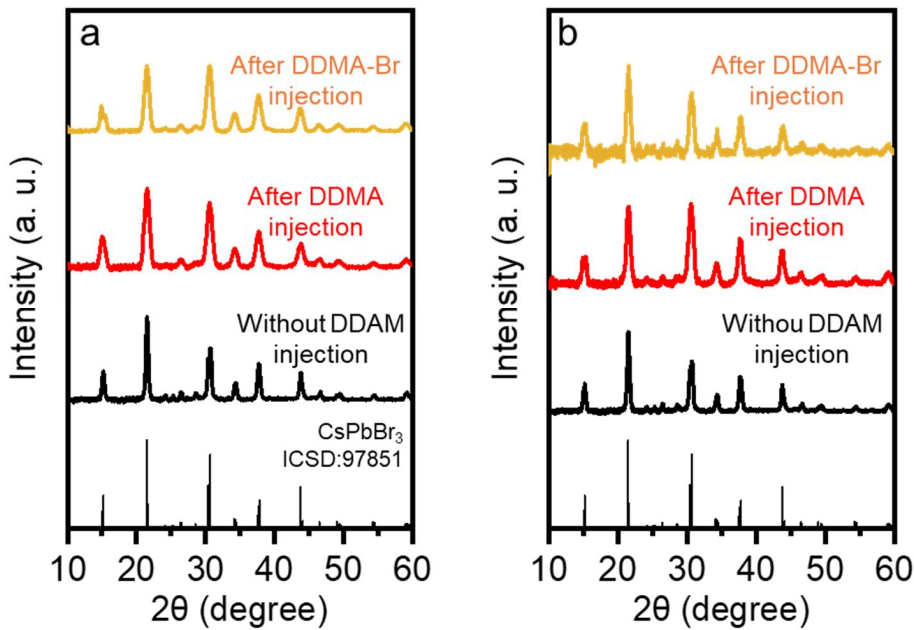

**Figure S21.** XRD patterns of CsPbBr$_3$ NCs synthesized employing either (a) Zn$^{2+}$ or (b) Sr$^{2+}$ as exogenous cations with oleic acid (black curves), and after adding either DDMA (red curves) or DDMA-Br (orange curves). When employing DDMA and DDMA-Br, the reactions where quenched 10 seconds after the injection of the amine/ammonium salt.

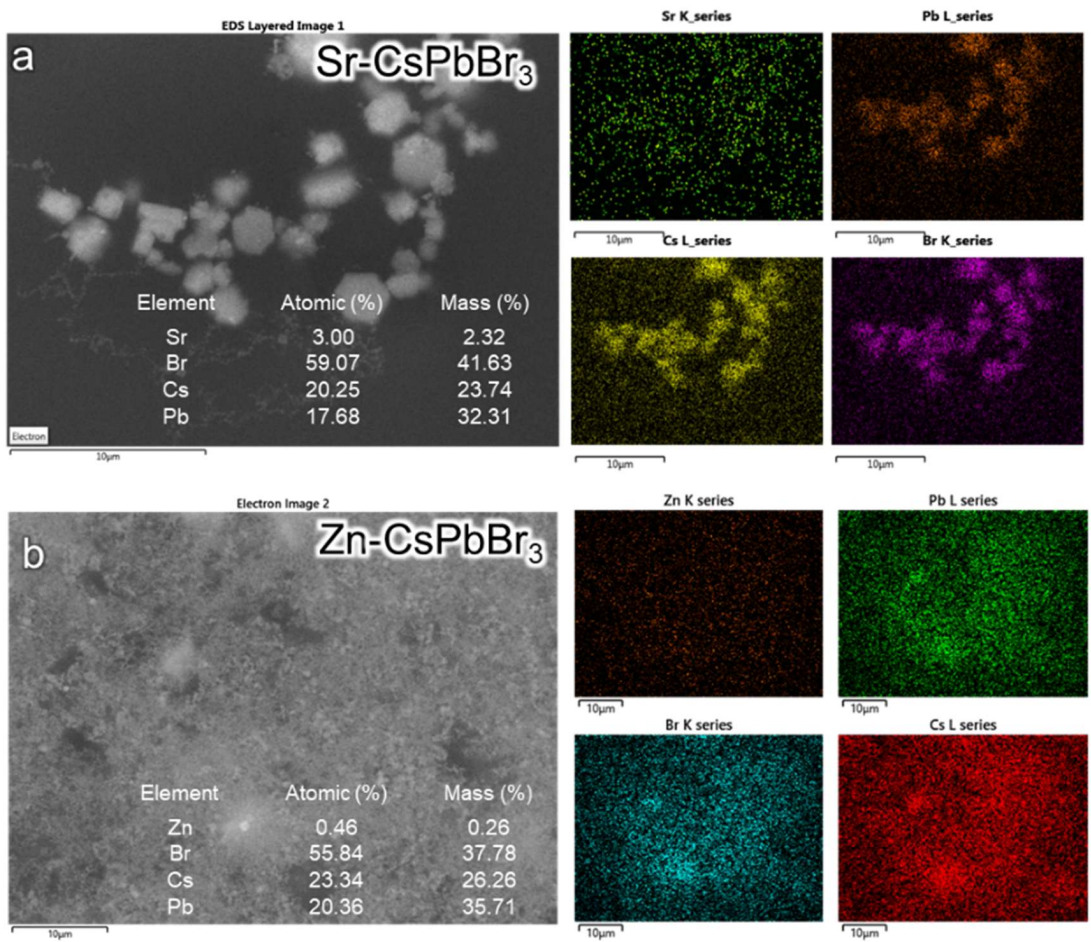

**Figure S22.** SEM-EDS analyses and corresponding element mapping of CsPbBr$_3$ NC samples made with either (a) Sr$^{2+}$ or (b) Zn$^{2+}$ and DDMA (the reactions were quenched 10 sec after the amine injection).

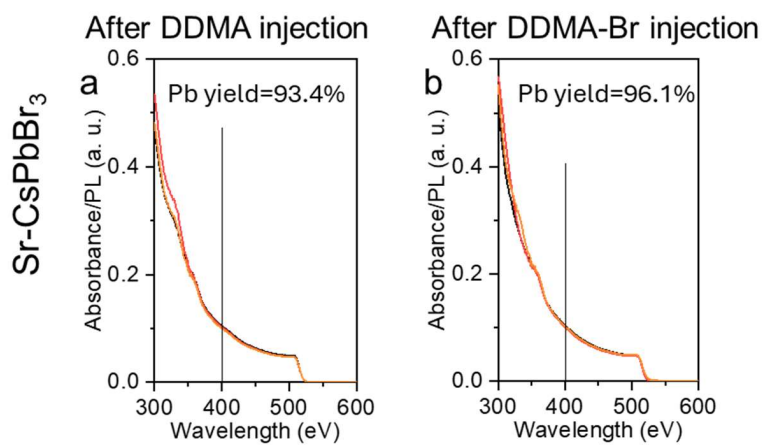

**Figure S23.** Estimation of the Pb yield by optical absorption spectroscopy of the crude solution of CsPbBr$_3$ NCs made employing Sr$^{2+}$ cations in the "minimal synthesis" approach with either (a) DDMA or (b) DDMA-Br. The reaction was quenched 10 sec after the addition of DDMA or DDMA-Br.

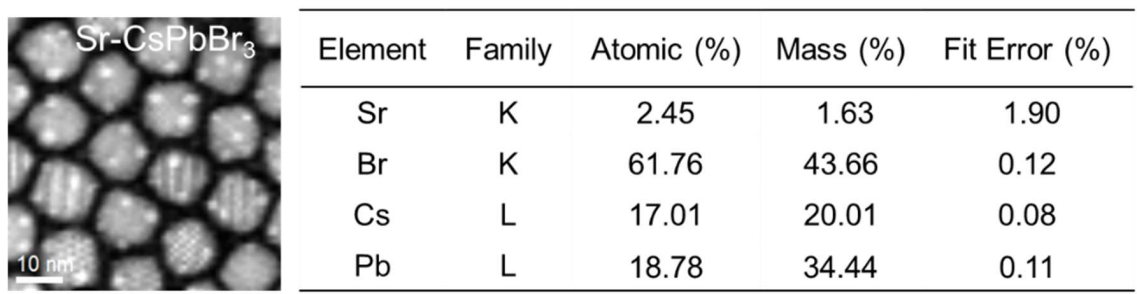

| Element | Family | Atomic (%) | Mass (%) | Fit Error (%) |
|---|---|---|---|---|
| Sr | K | 2.45 | 1.63 | 1.90 |
| Br | K | 61.76 | 43.66 | 0.12 |
| Cs | L | 17.01 | 20.01 | 0.08 |
| Pb | L | 18.78 | 34.44 | 0.11 |

**Figure S24.** HAADF-STEM and EDS analyses of CsPbBr$_3$ NCs made with Sr$^{2+}$ and DDMA (the reactions were quenched 10 sec after the amine injection).

**Results of the reactions in which the reaction time was prolonged to 20min**

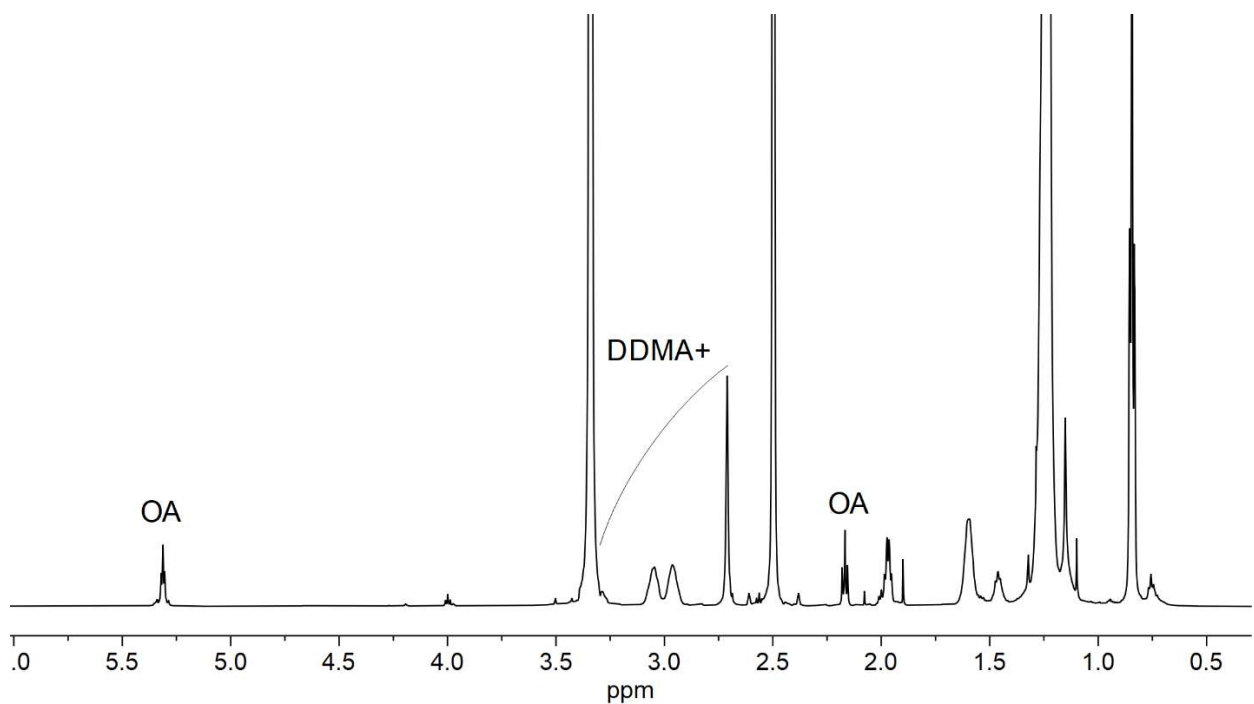

**Figure S25.** $^1$H NMR quantitative spectrum of CsPbBr$_3$ NCs (made with Zn$^{2+}$ and DDMA-Br, reaction quenched after 20 min) dissolved in DMSO-D.

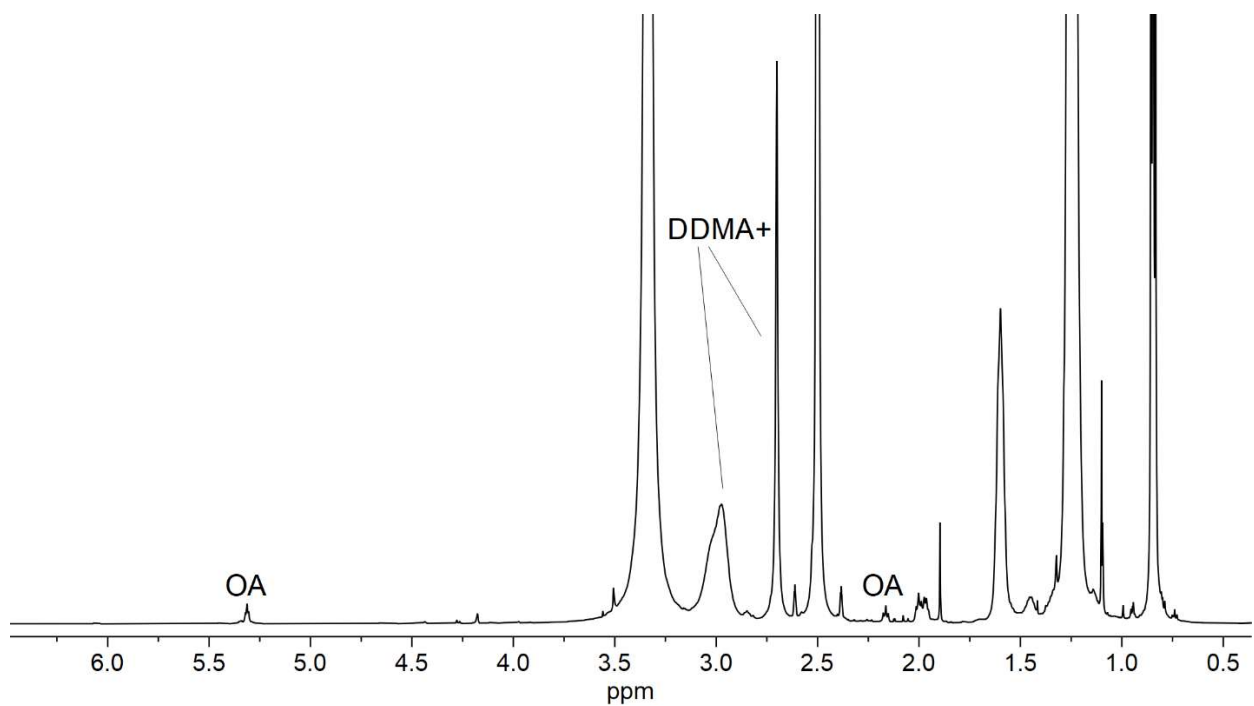

**Figure S26.** $^1$H NMR quantitative spectrum of CsPbBr$_3$ NCs (made with Sr$^{2+}$ and DDMA, reaction quenched after 20 min) dissolved in DMSO-D.

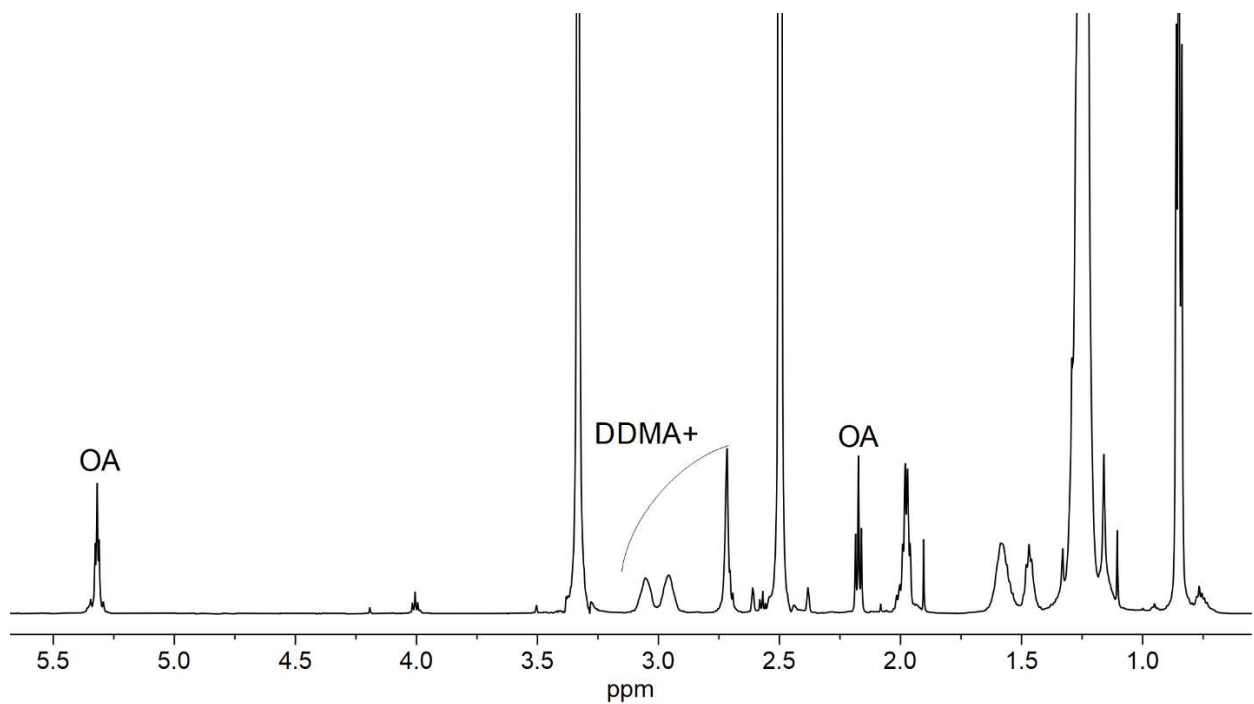

**Figure S27.** $^1$H NMR quantitative spectrum of CsPbBr$_3$ NCs (made with Sr$^{2+}$ and DDMA-Br, reaction quenched after 20 min) dissolved in DMSO-D.

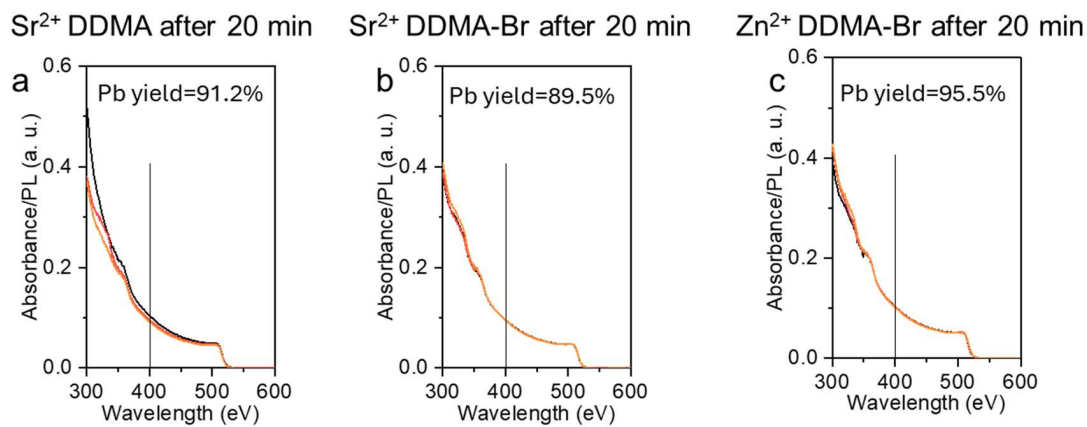

**Figure S28.** Estimation of the Pb yield by optical absorption spectroscopy of the crude solution of CsPbBr$_3$ NCs made employing Sr$^{2+}$ cations in the "minimal synthesis" approach with either (a) DDMA or (b) DDMA-Br and by quenching the reaction 20min after the amine/ammonium injection. (c) Estimation of the Pb yield by optical absorption spectroscopy of the crude solution of CsPbBr$_3$ NCs made employing Zn$^{2+}$ cations in the "minimal synthesis" approach with DDMA-Br and by quenching the reaction 20min after the ammonium injection.

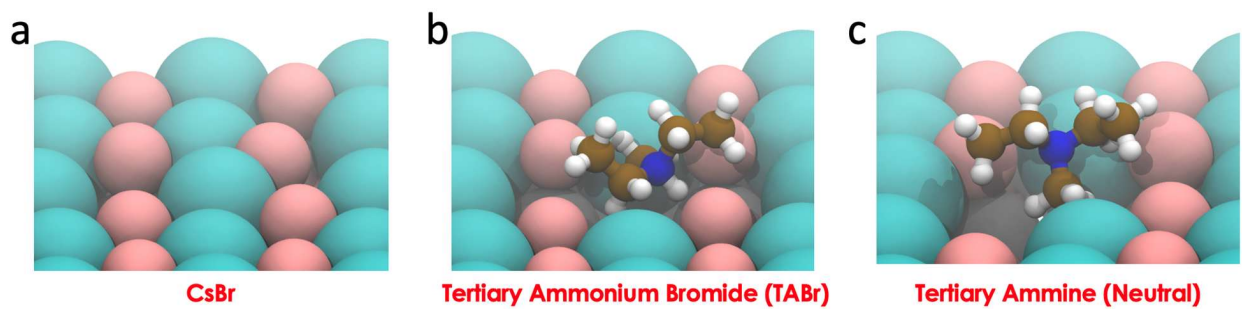

**Figure S29.** Sketches of the binding features for (a) a CsBr ion pair, (b) a tertiary ammonium-Br ion pair and (c) a neutral tertiary amine at the (001) facet of 2.4 nm-sided cubic CsPbBr$_3$ NC model after relaxation at the DFT/PBE/DZVP level of theory. The following color code is employed: Cs: cyan; Pb: grey; Br: pink; C: brown; N: blue; H: white.

**Table S1.** Size and M/Pb ratio of the NC samples reported in Figure 7 of the main text. The size was calculated from the TEM pictures of the samples while the M/Pb ratio was measured via ICP-OES.

| Sample | Size (nm) | M/Pb ratio |
|---|---|---|
| Hg-CsPbBr$_3$ | 12.4 ± 0.6 | 0.07 |
| Cu-CsPbBr$_3$ | 14.7 ± 0.7 | 0.19* |
| Sn-CsPbBr$_3$ | 13.4 ± 0.7 | 0.10 |
| Pb-CsPbBr$_3$ | 12.1 ± 0.5 | / |
| In-CsPbBr$_3$ | 9.5 ± 0.5 | 0.01 |
| Al-CsPbBr$_3$ | 10.3 ± 0.5 | 0.02 |
| Ni-CsPbBr$_3$ | 11.8 ± 0.7 | 0.02 |
| Mg-CsPbBr$_3$ | 12.0 ± 0.6 | 0.01 |
| Ca-CsPbBr$_3$ | 11.6 ± 0.5 | 0.01 |
| Ba-CsPbBr$_3$ | 11.7 ± 0.5 | 0.02 |
| Mn-CsPbBr$_3$ | 12.6 ± 0.5 | 0.03 |
| Na-CsPbBr$_3$ | 9.9 ± 0.5 | 0.07 |

* In this case the NCs were difficult to purify. However, both the optical properties and XRD pattern of Cu-based NCs were consistent with those of "pure" CsPbBr$_3$ NCs, strongly suggesting that the actual incorporation of Cu cations is very low.

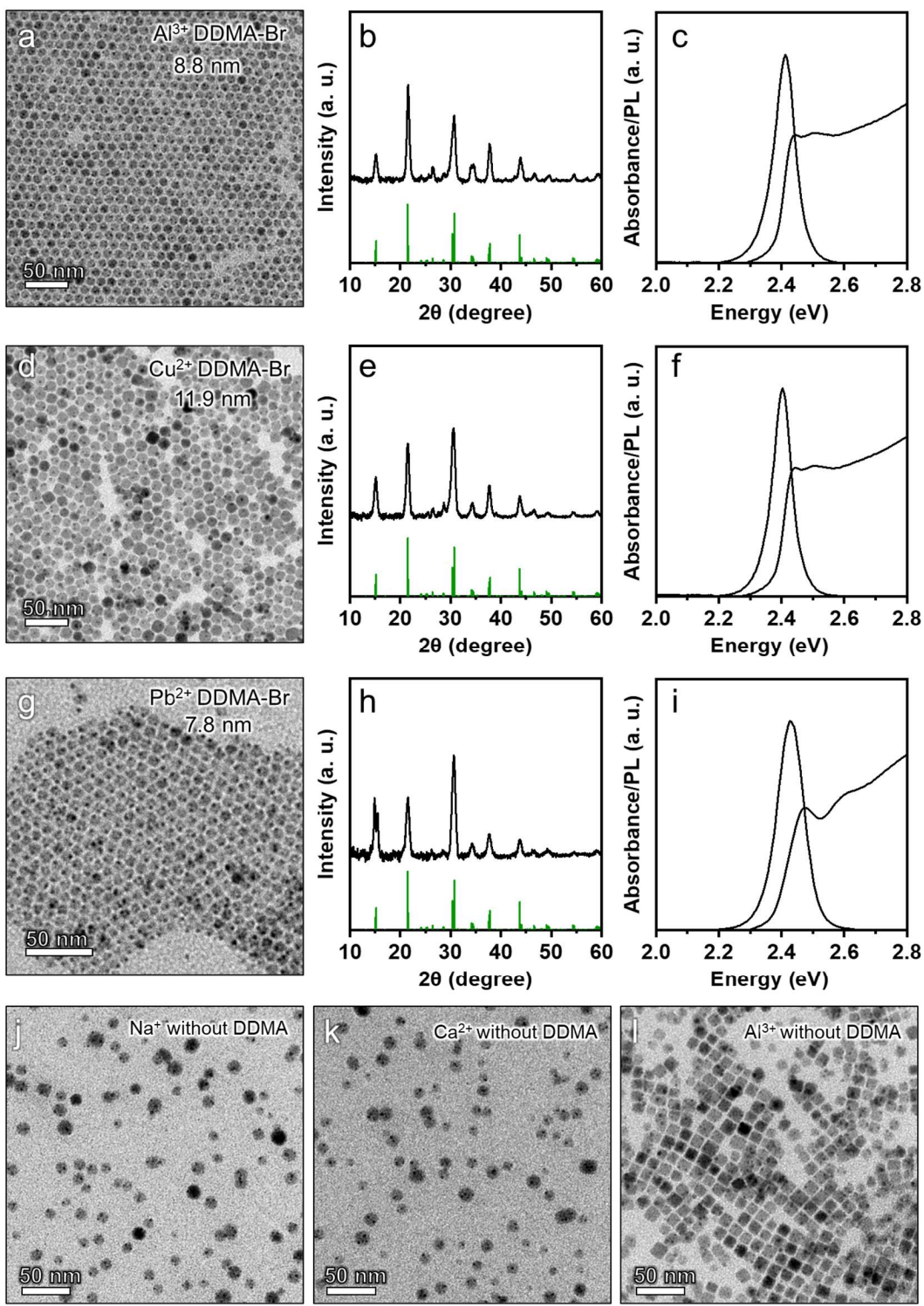

**Figure S30.** (a, d, g) TEM images, (b, e, h) XRD patterns and (c, f, i) optical absorption and PL spectra of the NC samples made using $Al^{3+}$, $Cu^{2+}$ and "$Pb^{2+}$" as exogenous cations and DDMA-Br (and quenching the reaction 10 sec after the addition of the ammonium bromide salt). In (b, e, h) the reflections of bulk $CsPbBr_3$ (ICSD number 97851) are represented by vertical green bars. TEM

images of CsPbBr$_3$ NCs obtained when using (j) Na$^+$, (k) Ca$^{2+}$ and (l) Al$^{3+}$ as exogenous cations without the addition of DDMA.

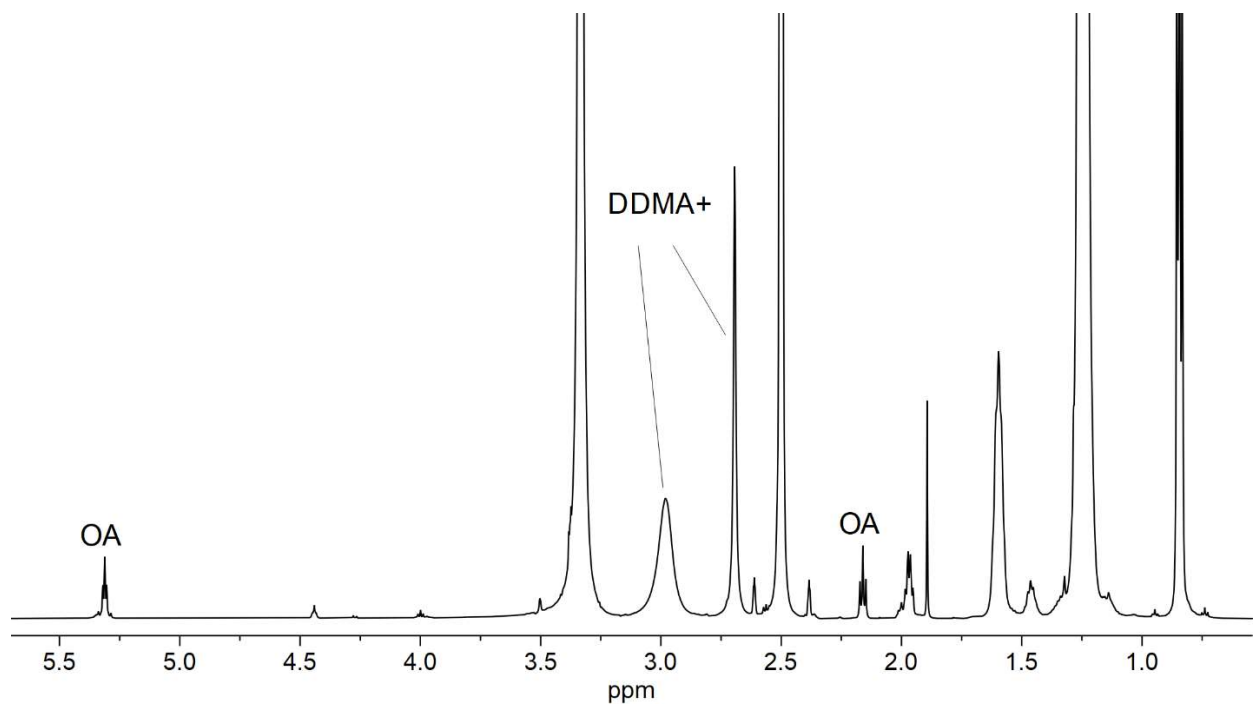

**Figure S31.** $^1$H NMR quantitative spectrum of CsPbBr$_3$ NCs (made with Ca$^{2+}$ and DDMA, reaction quenched after 10 sec) dissolved in DMSO-D.

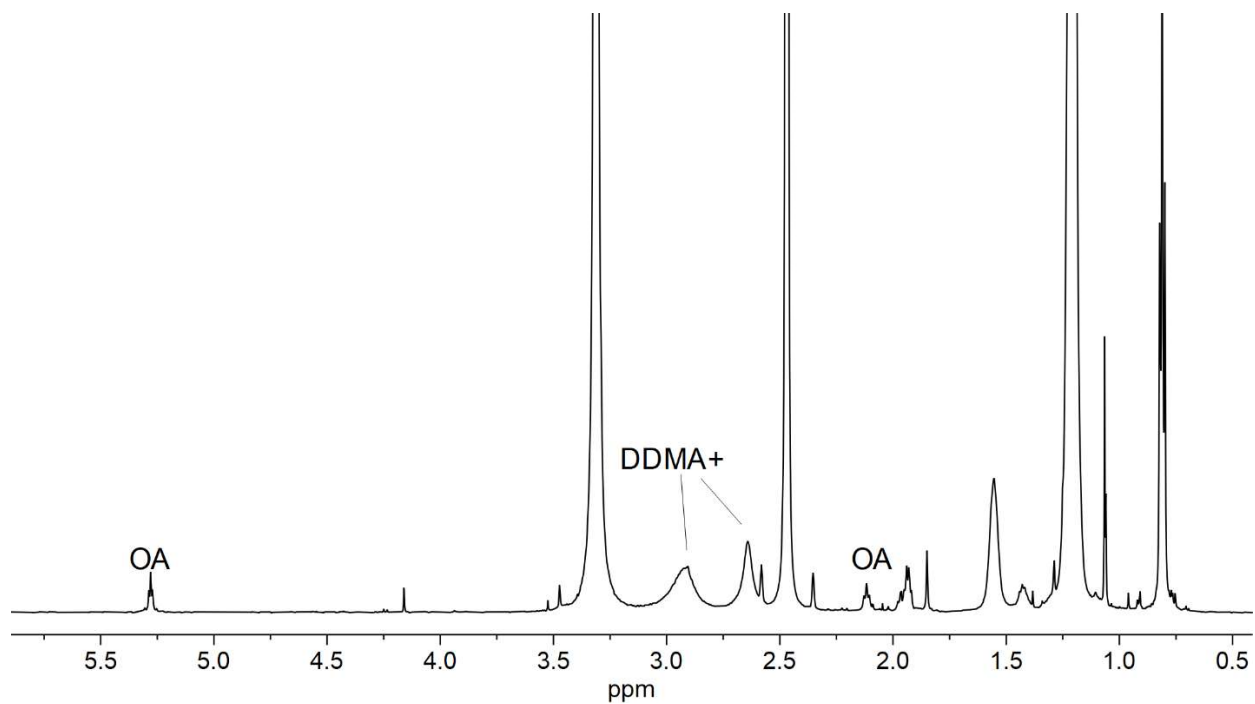

**Figure S32.** $^1$H NMR quantitative spectrum of CsPbBr$_3$ NCs (made with Na$^+$ and DDMA, reaction quenched after 10 sec) dissolved in DMSO-D.

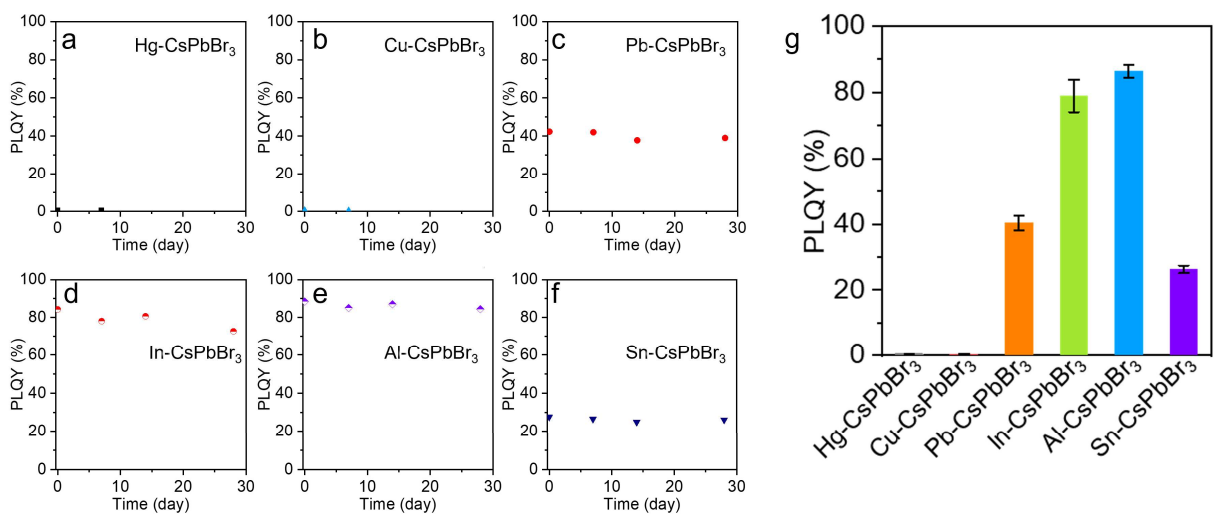

**Figure S33.** (a-f) PLQY values measured for CsPbBr$_3$ NCs made with (a) Hg$^{2+}$, (b) Cu$^{2+}$, (c) Pb$^{2+}$, (d) In$^{3+}$, (e) Al$^{3+}$, (f) Sn$^{4+}$ and DDMA (the reactions were quenched 10 sec after the injection of the amine), over a time span of one month. (g) PLQY values of the NC samples obtained with the different M cations.

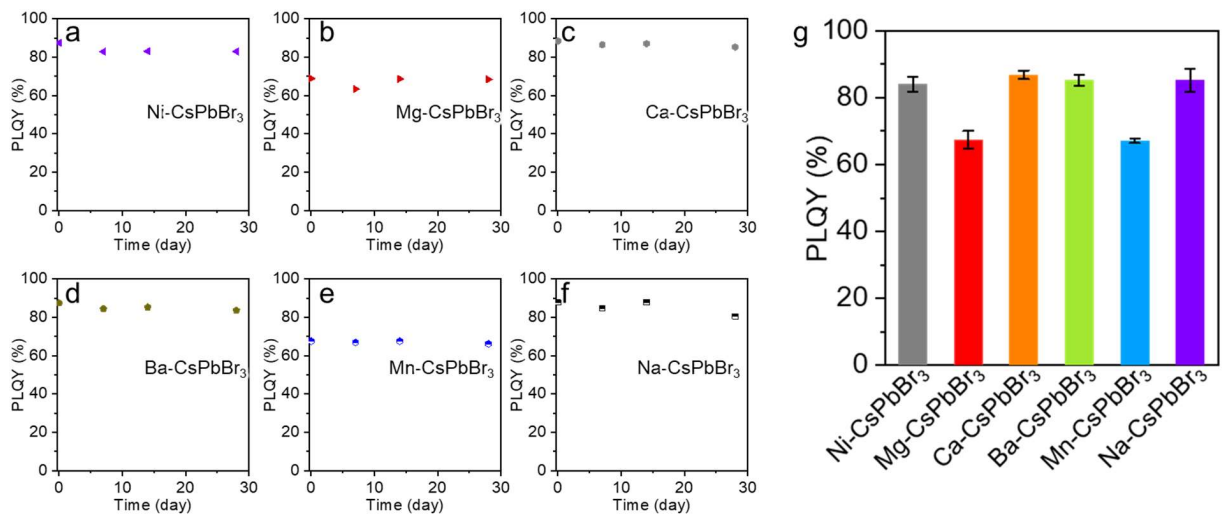

**Figure S34.** (a-f) PLQY values measured for CsPbBr$_3$ NCs made (a) Ni$^{2+}$, (b) Mg$^{2+}$, (c) Ca$^{2+}$, (d) Ba$^{2+}$, (e) Mn$^{2+}$, (f) Na$^+$ and DDMA (the reactions were quenched 10 sec after the injection of the amine), over a time span of one month. (g) PLQY values of the NC samples obtained with the different M cations.

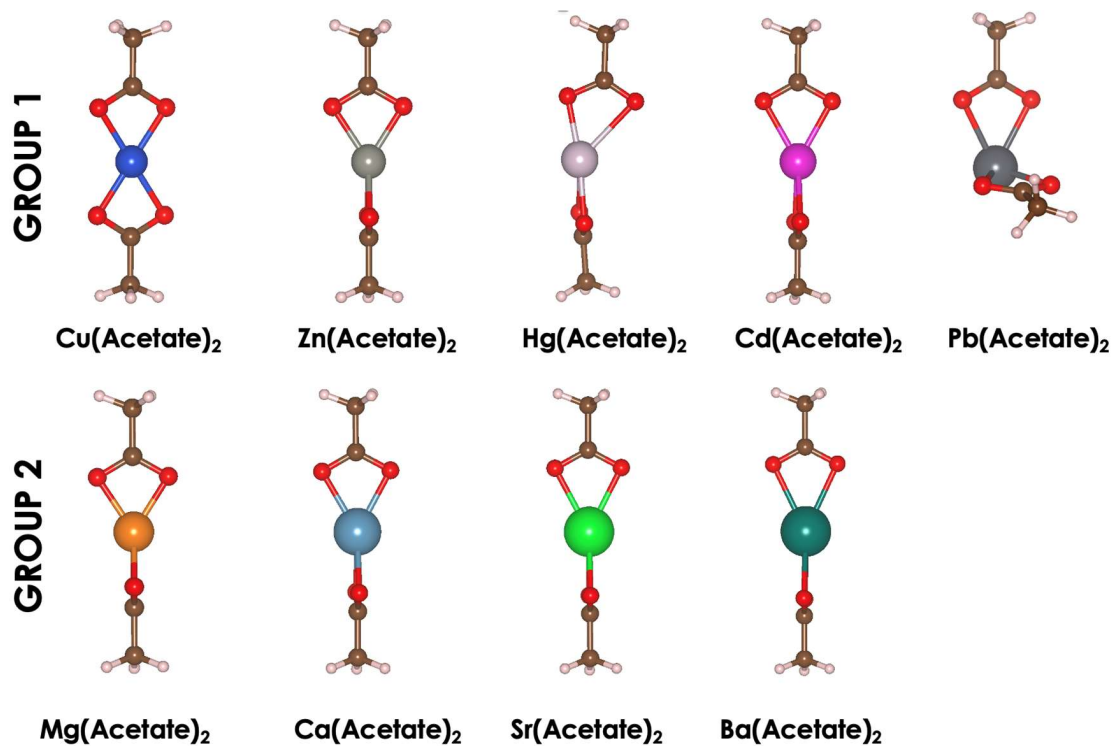

**Figure S35.** Ball and sticks representation of the M(acetate)$_x$ molecular complexes structures relaxed at the DFT/PBE/DZVP level of theory.

**References**

(1) Maes, J.; Balcaen, L.; Drijvers, E.; Zhao, Q.; De Roo, J.; Vantomme, A.; Vanhaecke, F.; Geiregat, P.; Hens, Z. Light Absorption Coefficient of CsPbBr$_3$ Perovskite Nanocrystals. *J. Phys. Chem. Lett.* **2018**, *9* (11), 3093−3097.
(2) Abraham, R. J.; Fisher, J.; Loftus, P. Introduction to NMR Spectroscopy. Wiley, **1988**.
(3) Montgomery, C. D. Factors Affecting Energy Barriers for Pyramidal Inversion in Amines and Phosphines: A Computational Chemistry Lab Exercise. *J. Chem. Educ.* **2013**, *90* (5), 661−664.